\documentclass[nofootinbib,prl,amsmath,amssymb,twocolumn,floatfix]{revtex4-2}
\usepackage{graphicx}
\usepackage{dcolumn}
\usepackage{siunitx}
\usepackage{bm}
\usepackage{color}
\newcommand{\bxi}{\boldsymbol{\xi}}

\newcommand{\bPsi}{\boldsymbol{\Psi}}

\DeclareMathOperator{\tr}{tr}
\usepackage{hyperref}
\hypersetup{
    linkcolor=blue,
    pdfpagemode=FullScreen,
    }
    
\newcommand\blfootnote[1]{%
  \begingroup
  \renewcommand\thefootnote{}\footnote{#1}%
  \addtocounter{footnote}{-1}%
  \endgroup
}

\begin{document}

\preprint{APS/123-QED}

\title{Elastic Orbital Angular Momentum}

\author{G.~J. Chaplain$^{1*}$, J.~M. De Ponti$^2$ and R.~V. Craster$^{3,4,5}$}
\affiliation{$^1$Electromagnetic and Acoustic Materials Group, Department of Physics and Astronomy, University of Exeter, Exeter EX4 4QL, United Kingdom} 
\affiliation{$^2$ Department of Civil and Environmental Engineering, Politecnico di Milano, Piazza Leonardo da Vinci, 32, 20133 Milano, Italy }
\affiliation{$^3$ Department of Mathematics, Imperial College London, 180 Queen's Gate, South Kensington, London SW7 2AZ \\
$^4$ Department of Mechanical Engineering, Imperial College London, London SW7 2AZ, UK \\
$^5$ UMI 2004 Abraham de Moivre-CNRS, Imperial College London, London SW7 2AZ, UK}
\email{g.j.chaplain@exeter.ac.uk}


\begin{abstract}
We identify that flexural guided elastic waves in elastic pipes carry a well-defined orbital angular momentum associated with the compressional dilatational potential. This enables the transfer of elastic orbital angular momentum, that we numerically demonstrate, through the coupling of the compressional potential in a pipe to the acoustic pressure field in a surrounding fluid in contact with the pipe.
\end{abstract}

\maketitle

\noindent\textit{Introduction.}| \blfootnote{This is the accepted version of \textit{G.J. Chaplain, J.M. De Ponti, and R.V. Craster, Elastic Orbital Angular Momentum, Phys. Rev. Lett. 128, 064301  (2022)}. The final publication is available at \textcolor{blue}{\href{https://doi.org/10.1103/PhysRevLett.128.064301}{https://doi.org/10.1103/PhysRevLett.128.064301}}}
Some thirty years ago, the seminal work of Allen et al. \cite{allen1992orbital} demonstrated that Laguerre--Gaussian (LG) laser modes carry a well-defined orbital angular momentum (OAM), per quanta of light, about the beam axis. Crucially they outlined how such OAM, related to the spatial distribution of the laser field \cite{jackson1999classical}, can be extracted and converted into a mechanical torque \cite{he1995direct} and that its existence arises physically due to the helical wave-front structure associated with a central phase singularity \cite{o2002intrinsic}. This differed from previous measurements of the torque exerted by the transfer of spin angular momentum associated with polarization \cite{beth1936mechanical,poynting1909wave}. These significant findings drove unabated interest in this previously neglected mechanical property of light \cite{barnett2001optical,barnett2010rotation,allen2016optical}, and have led to a renaissance in optical tweezers \cite{garces2003observation,yao2011orbital,willner2015optical,bliokh2015transverse,barnett2017optical,padgett2017orbital,chen2019orbital}.

Perhaps the most distinct classical wave system from electromagnetism is elasticity; elastic materials are governed by constitutive relations that invoke a rank 4 stiffness tensor, and even in their simplest isotropic form they support two  elastic waves (compression and polarised-shear) that travel within the bulk at distinct wave speeds; these become inherently coupled upon reflection from a surface. Mirroring the timeline of research in optical OAM, only recently has the intrinsic spin of elastic waves been studied \cite{long2018intrinsic}, with elastic OAM being largely neglected - it has only been considered in association with the phase of coupled waveguides \cite{deymier2018elastic}, or presented canonically in conjunction with the energy-momentum tensor for elasticity, the Eshelby tensor \cite{eshelby1951force,eshelby1975elastic,thielheim1967note,lazar2007eshelby}.

\begin{figure}[h]
    \centering
    \includegraphics[width = 0.5\textwidth]{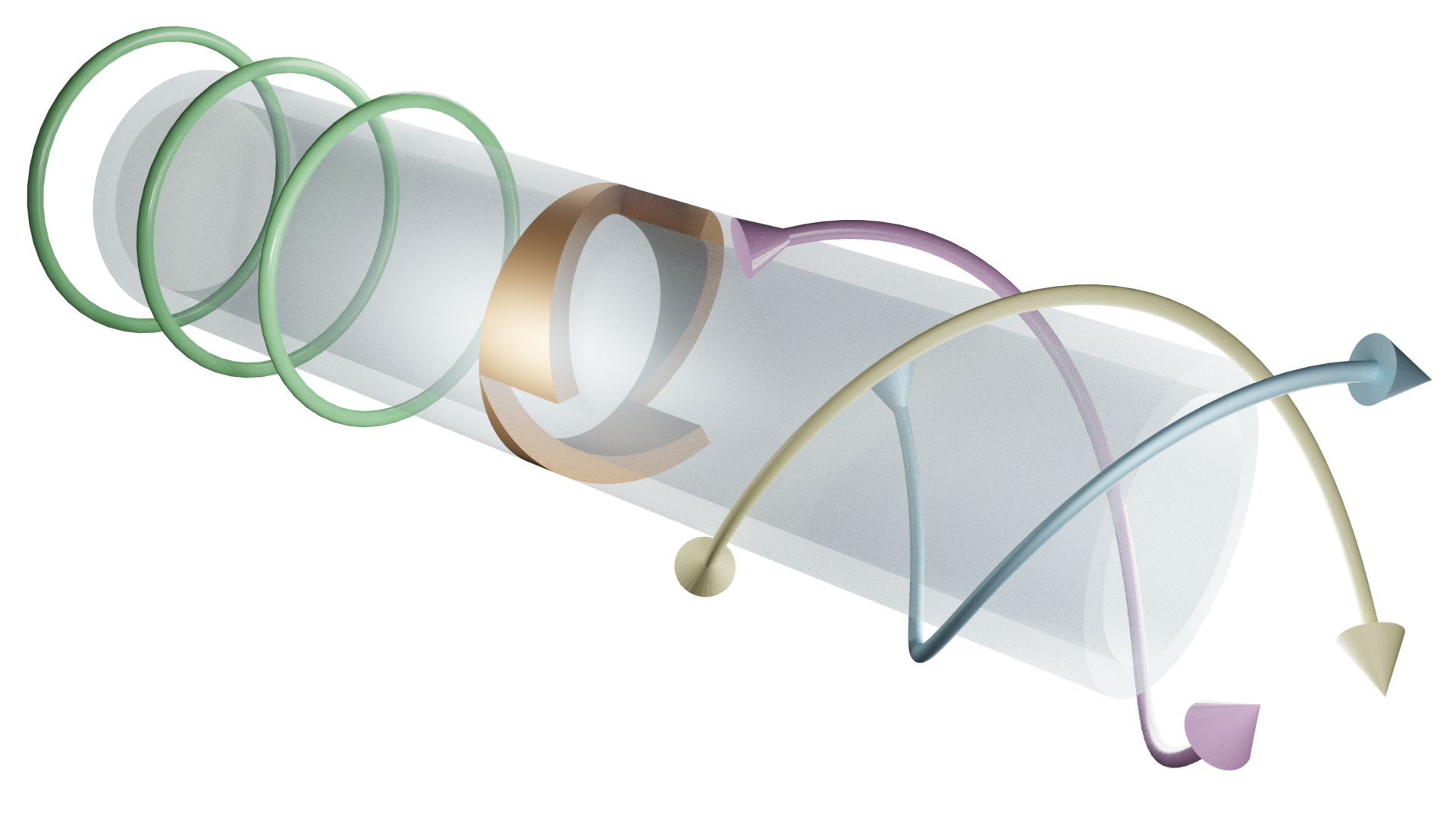}
    \caption{Schematic of an elastic spiral phase pipe (copper region) in a hollow elastic pipe (transparent region). Purely longitudinal waves, e.g. $L(0,2)$ modes (circular phase fronts), are mode converted into flexural $F(3,n)$ waves (helical phase fronts).}
    \label{fig:esp}
\end{figure}

In this letter this disparity is addressed. We focus entirely on the OAM of elastic waves with inclined phase fronts, demonstrating that it is the scalar dilatational potential which carries a well-defined elastic OAM. The natural setting for such guided waves is along hollow elastic pipes. We consider flexural modes along pipes, leveraging the fact they can be excited using an elastic analogue to the spiral phase plate (Fig.~\ref{fig:esp}), and show that the transfer of elastic OAM is possible in fluid-solid coupled systems, providing motivation towards applications for acoustic tweezers, microfluidic devices and non-destructive evaluation. 

To unequivocally show that elastic OAM is carried by mechanical waves in pipes, we first outline the form of the canonical angular momentum density by its relation to mechanical energy flux. We derive here, from first principles, this relation from the Eshelby tensor \cite{eshelby1975elastic}. 

\noindent\textit{OAM in elasticity.}|
Waves in an isotropic, homogeneous linear elastic material obey the dynamic Navier--Cauchy equation \cite{landau1959course}
\begin{equation}
    \mu\partial_{j}\partial_{j}\xi_{i} + (\lambda + \mu)\partial_{j}\partial_{i}\xi_{i} = \rho\ddot{\xi_{i}}
    \label{eq:elastod}
\end{equation}
with $\xi_{i}$ the displacement and $\ddot{\xi_{i}}$ its double time derivative. Lam\'{e}'s first and second parameters are denoted $\lambda$, $\mu$ respectively. In this coordinate-free index notation we adopt the Einstein summation convention throughout. The linear constitutive law governing such a material is
\begin{align}
\begin{split}
    \sigma_{ij} &= C_{ijkl}\varepsilon_{kl} = \lambda\delta_{ij}\varepsilon_{kk} + 2\mu\varepsilon_{ij},
    \end{split}
\end{align}
where $\sigma_{ij}$ is the stress tensor, $C_{ijkl}$ is the stiffness tensor and $\varepsilon_{ij} \equiv \frac{1}{2}(\xi_{i,j} + \xi_{j,i})$ is the strain tensor (comma notation denotes partial differentiation). The elastodynamic equations \eqref{eq:elastod} are, of course, recovered by the Euler--Lagrange equations that dictate the vanishing of the variational derivative 
\begin{equation}
    \frac{\delta\mathcal{L}}{\delta\xi_{j}} \equiv \frac{\partial\mathcal{L}}{\partial\xi_{j}} - \frac{\partial}{\partial t}\left(\frac{\partial\mathcal{L}}{\partial\dot{\xi_j}}\right) = 0,
\end{equation}
with $\mathcal{L}$ being the Lagrangian density for elastic waves given by
\begin{equation}
    \mathcal{L} = \frac{1}{2}\rho\dot{\xi_i}\dot{\xi_i} - \frac{1}{2}C_{ijkl}\xi_{i;j}\xi_{k;l},
\end{equation}
where semicolon notation denotes covariant differentiation and $\mathcal{L} = \mathcal{L}(\xi_{i},\xi_{i,j},\boldsymbol{x},t)$ with $\boldsymbol{x}$ the position vector. The Eshelby tensor results from the canonical procedure for constructing stress-energy tensors, following Noether's theorem, and is given as \cite{noether1971invariant}
\begin{equation}
    T_{lj} = \mathcal{L}\delta_{lj} -\frac{\partial\mathcal{L}}{\partial\xi_{i,j}}\xi_{i,l}.
\end{equation}

From this the energy density, $U = T_{00}$, and flux, $F_{j} = T_{0j}$, of the elastic waves can be constructed:
\begin{align}
    \begin{split}
        U &= \frac{\partial\mathcal{L}}{\partial\dot{\xi_{i}}}\dot{\xi_{i}} - \mathcal{L} = \frac{1}{2}\rho\dot{\xi_{i}}\dot{\xi_{i}} + \frac{1}{2}C_{ijkl}\xi_{i;j}\xi_{k;l} \\
        F_{j} &= \frac{\partial\mathcal{L}}{\partial\xi_{i;j}}\dot{\xi_{i}} = -C_{ijkl}\dot{\xi_i}\xi_{k;l}.
    \end{split}
\end{align}
Thus we have arrived at the mechanical analogue of the Poynting vector through the mechanical energy flux, $F_j$. Herein we assume time harmonicity such that $\dot{\xi_{k}} = -i\omega\xi_{k}$ with $\omega$ being the radian frequency. Therefore the time-averaged complex mechanical energy flux density, which can be considered the Poynting vector density of elastic waves \cite{long2018intrinsic}, is written as
\begin{equation}
    F_{j} = -\frac{1}{2}\mathfrak{Re}\left(\sigma_{ji}\dot{\xi_{i}}^{*} \right) = -\frac{\omega}{2}\mathfrak{Im}\left(\sigma_{ji} \xi_{i}^{*} \right),
    \label{eq:energyflux}
\end{equation}
where $^{*}$ denotes complex conjugation. The integral of this quantity, as in electromagnetism, is thus interpreted as the linear momentum density. 

The flux of the corresponding angular momentum density is defined by the rank 3 tensor $M_{ijk} = x_{i}T_{jk} - x_{j}T_{ik}$ \cite{barnett2001optical}. The antisymmetric pseudo-tensor of rank 2, $M_{ij0} = \epsilon_{ilm}x_{l}T_{mj}$ has spatial components, i.e. the pseudo-vector $M_{i} = 1/2\epsilon_{ijk}M_{jk0}$ \cite{jackson1999classical} that are then the familiar angular momentum density of the form, in vector notation, $\boldsymbol{M} = \boldsymbol{r}\times\langle\boldsymbol{p}\rangle$ with $\langle\boldsymbol{p}\rangle$ the time-averaged linear momentum density. For convenience we now switch from index notation to coordinate dependent vector notation and explicitly consider cylindrical polar coordinates. 

Analogous to the treatment of electromagnetic waves in \cite{allen1992orbital} we now consider the elastic angular momentum density as
\begin{equation}
    \boldsymbol{M} = -\frac{\omega}{2}\mathfrak{Im}[\boldsymbol{r} \times \left(\underline{\underline{{\sigma}}} \bm{\cdot} \bxi^{*}\right)],
    \label{eq:angmom}
\end{equation}
highlighting the tensorial nature of the stress tensor with a double-underline, $\underline{\underline{{\sigma}}}$, such that the total angular momentum is then
\begin{equation}
    \boldsymbol{\mathcal{J}} = -\frac{\omega}{2}\mathfrak{Im}\int\boldsymbol{r} \times \left(\underline{\underline{{\sigma}}} \bm{\cdot} \bxi^{*}\right)d\mathbf{r}.
\end{equation}
The complex displacement field $\bxi$ is separated into longitudinal and transverse components via Helmholtz decomposition. These are written respectively in terms of the curl-less dilatational scalar potential, $\Phi$ (analogous to the scalar potential of the LG beams in optics, see supplemental material \cite{SM}), and the divergence-less equivoluminal vector shear potential, $\bPsi$, such that
\begin{align}
\begin{split}
    \bxi &= \bxi_{L} + \bxi_{T} = \nabla\Phi + \nabla\times\bPsi 
    \end{split}
\end{align}
where $\bxi_{L}$ and $\bxi_{T}$ denote the longitudinal and transverse parts respectively. Using this, the elastodynamic equations \eqref{eq:elastod} reduce to two wave equations for compressional and shear waves:
\begin{align}
\begin{split}
    \nabla^2\Phi &= c_p^{-2}\ddot{\Phi}, \quad c_p = \sqrt{\frac{\lambda + 2\mu}{\rho}}, \\
    \nabla^2\bPsi &= c_s^{-2}\ddot{\bPsi}, \quad c_s = \sqrt{\frac{\mu}{\rho}},
\end{split}
\label{eq:waveeqns}
\end{align}
with $c_p$ and $c_s$ being the compressional and shear bulk wavespeeds respectively.  

The angular momentum density can therefore be re-written in terms of its spin, orbit and `additional' components \cite{lazar2007eshelby}, each with individual contributions from the shear and compressional potentials. Long et al. \cite{long2018intrinsic} identify this additional component as hybrid orbital and spin Poynting densities.

The energy flux density \eqref{eq:energyflux} can then be split into orbital components which take the form
\begin{align}
    \begin{split}
        \boldsymbol{p}_{L}^{o} &= \frac{\omega\rho}{2}c_p^2\mathfrak{Im}\left[(\bxi_{L}^{*}\bm{\cdot}\nabla)\bxi_{L} \right], \\
        \boldsymbol{p}_{T}^{o} &= \frac{\omega\rho}{2}c_s^2\mathfrak{Im}\left[(\bxi_{T}^{*}\bm{\cdot}\nabla)\bxi_{T} \right], \\
        \boldsymbol{p}_{H}^{o} &= \frac{\omega\rho}{2}c_p^2\mathfrak{Im}\left[(\bxi_{T}^{*}\bm{\cdot}\nabla)\bxi_{L} \right] + \frac{\omega\rho}{2}c_s^2\mathfrak{Im}\left[(\bxi_{L}^{*}\bm{\cdot}\nabla)\bxi_{T} \right].
    \end{split}
    \label{eq:pol}
\end{align}
with the subscripts $L,T,H$ corresponding to the  longitudinal, transverse (shear) and hybrid parts respectively. We now prove that, for displacement fields with inclined phase fronts, the longitudinal part of the wave field $\bxi_{L}$, associated with compressional motion, carries a well-defined OAM.

Flexural waves in pipes serve as exemplar mode shapes capable of carrying elastic OAM. We consider elastic waves propagating along an infinitely long, hollow elastic cylinder with axis oriented in the $z$-direction of inner radius $r_a$ and outer radius $r_b$. The first general solution for these guided harmonic waves was derived by Gazis \cite{gazis1959a,gazis1959b}, who showed there are three families of modes: longitudinal, torsional and flexural. The naming convention for such modes classifies these as $L(m,n)$, $T(m,n)$ and $F(m,n)$ respectively \cite{silk1979propagation}. Here $m$ denotes the circumferential order, or azimuthal index, with $n$ the group order. The mode shapes for which $m = 0$ are axisymmetric i.e. their angular profile is constant. We consider non-axisymmetric flexural modes $F(m > 0,n)$ whose mode shapes vary sinusoidally in the circumferential direction.
\begin{figure}[t]
    \centering
    \includegraphics[width = 0.48\textwidth]{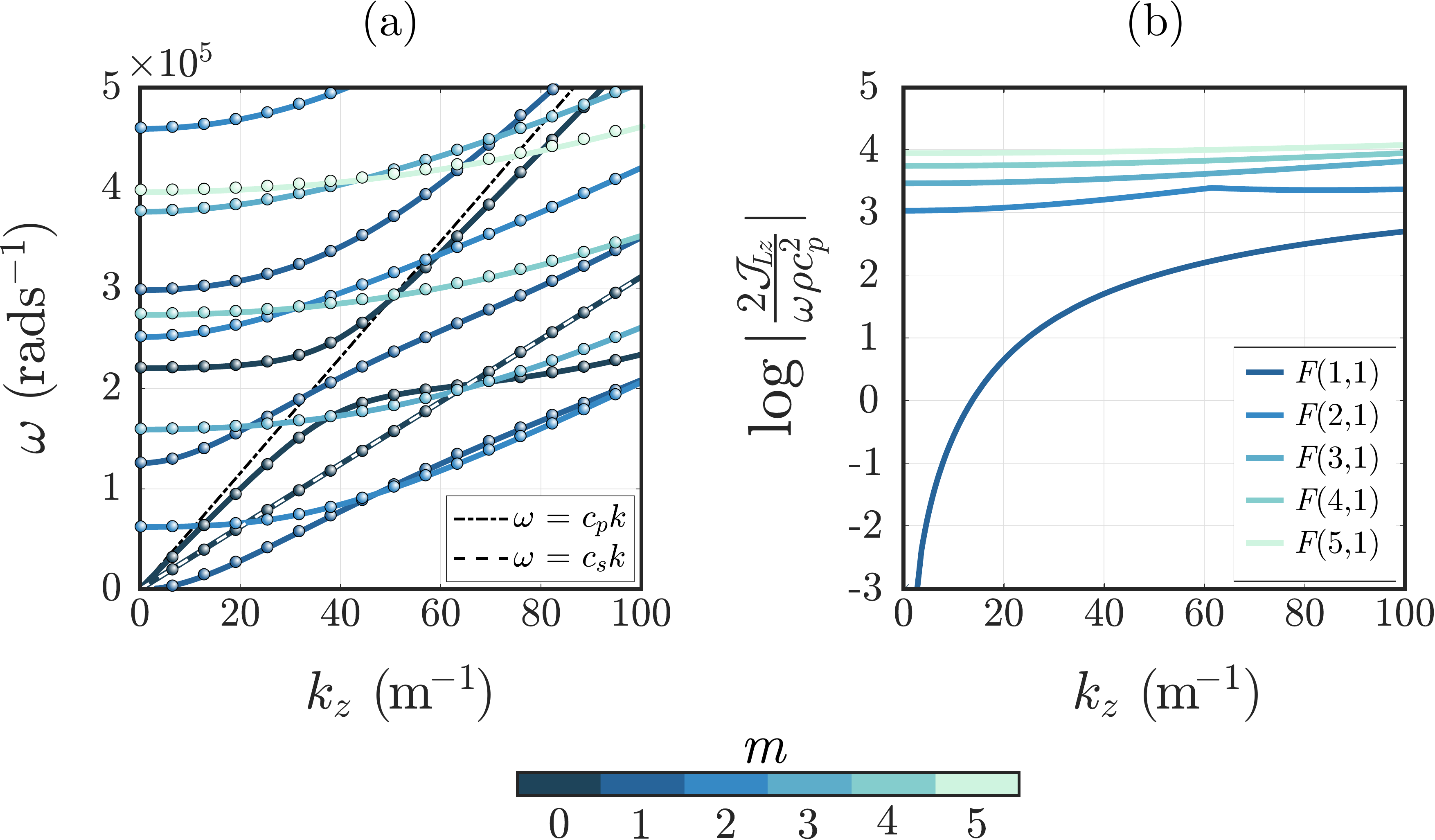}
    \caption{(a) Dispersion curves of guided waves in an aluminium pipe of inner diameter $40$~mm and thickness $10$~mm. Solid lines obtained by Spectral Collocation with points resulting from Finite Element computations.  (b) Numerical evaluation (details reported in the supplemental material \cite{SM}) of the well-defined elastic OAM along the pipe axis, $\mathcal{J}_{Lz}$, associated with the dilatational potential for the lowest curves of the flexural modes $F(m,1)$.}
    \label{fig:jzdisp}
\end{figure}
Following the ansatz of Gazis, we leverage the cylindrical symmetry of the pipe and writing the coordinate system as $\left(\boldsymbol{\hat{r}},\boldsymbol{\hat{\theta}},\boldsymbol{\hat{z}} \right)$ pose the form of the scalar dilatational potential as
\begin{equation}
     \Phi = \phi(r)\exp \left[i(m\theta + k_{z}z - \omega t)\right].
     \label{eq:ansatz}
\end{equation}
The compressional displacement field then has the form 
\begin{align}
\begin{split}
\bxi_{L} =  \left(\phi^{\prime},\frac{im\phi}{r},ik\phi \right)\exp\left[i(m\theta + k_{z}z - \omega t)\right], 
\end{split}
\end{align}
with the prime notation denoting partial differentiation with respect to $r$. After substitution into \eqref{eq:waveeqns} the radial profile $\phi(r)$ is solved by a linear combination of Bessel's functions, each with a complex amplitude. These coefficients are solved for by employing the infinitely long cylinder gauge, $\nabla \bm{\cdot} \bPsi = 0$, and traction free boundary conditions on the inner and outer radii $\sigma_{rr} = \sigma_{r\theta} = \sigma_{rz} = 0 \big\vert_{r_{a},r_{b}}$ (see supplemental material \cite{SM}). As such the guided modes in elastic pipes can be thought of as Bessel `beams' in the sense that the radial distributions satisfy Bessel's equation.

The contribution of $\boldsymbol{p}^{o}_{L}$ to the OAM density along the pipe axis $\boldsymbol{M}\bm{\cdot}\hat{\bm{z}}$ is defined as $\boldsymbol{M}_{L}^{o}\bm{\cdot}\hat{\bm{z}} = r\boldsymbol{p}_{L}^{o}\bm{\cdot}\hat{\bm{\theta}}$, where $\boldsymbol{p}_{L}^{o}\bm{\cdot}\hat{\bm{\theta}}$ is the azimuthal component of the orbital, longitudinal part of the linear momentum density. To evaluate this quantity we are required to evaluate the advective terms that arise in \eqref{eq:pol} due to the variation of the Lagrangian basis vectors as the
body deforms, highlighting its extrinsic nature; for an elastic deformation $\bxi$, ${\boldsymbol{\alpha}}$ is the Lagrangian position
vector ${\boldsymbol{\alpha}} = \boldsymbol{\beta} - \bxi$ with $\boldsymbol{\beta}$ the Eulerian position vector after the deformation. 

Calculation of the elastic OAM density along the pipe axis yields
\begin{widetext}
\begin{equation}
    \boldsymbol{M}_{L}^{o}\bm{\cdot}\hat{\bm{z}} = m\left\{\frac{\omega\rho c_p^2}{2}\left[|\phi^{\prime}|^2 +\left(\frac{1}{r}\left(r\phi^{\prime\prime} + \phi^{\prime}\right)\right)\phi^{*} + \frac{\omega^2}{c_p^2}|\phi|^2 -\frac{2}{r}\mathfrak{Re}\left(\phi\phi^{*\prime}\right) \right]\right\}.
\end{equation}
\end{widetext}
\begin{figure*}[ht!]
    \centering
    \includegraphics[width = \textwidth]{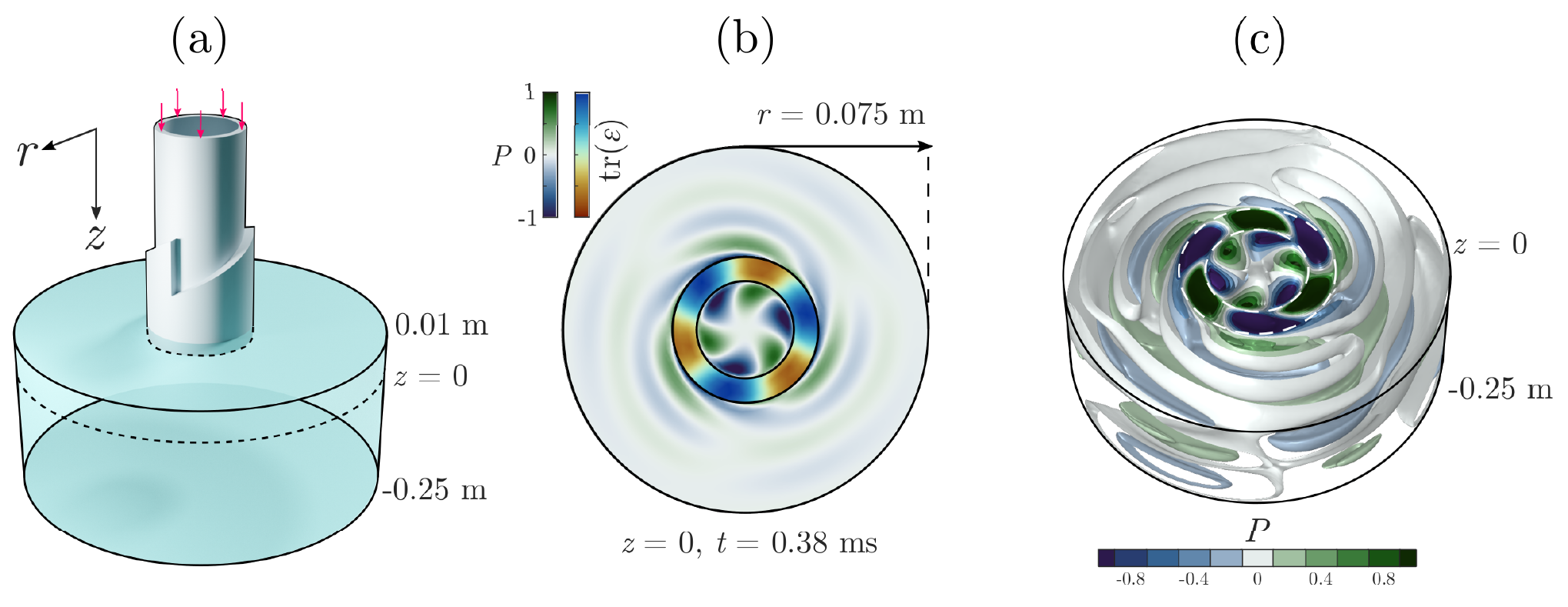}
    \caption{Finite Element time domain simulation of OAM transfer: (a) Schematic of simulation domain, with pipe partially submerged $1$ cm in water, surrounded by air. Arrows show axisymmetric longitudinal excitation position, with absorbing boundaries on exterior fluid walls. (b) Normalised compressional field in pipe (trace of the strain tensor, $\tr(\varepsilon)$) and pressure field ($P$) in the fluid at the end of the pipe ($z = 0$). (c) Isosurfaces of fluid pressure in region below the pipe showing spiraling acoustic waves, at same time instance as in (b). Full details shown in supplemental material, along with frequency domain corroborations \cite{SM}.}
    \label{fig:Sim}
\end{figure*}
In general $\phi(r)$ may be arbitrary and as such this result holds for all compressional wave fields with an azimuthally dependent profile, i.e. with inclined phase fronts, as is the case in electromagnetism \cite{o2002intrinsic}. Therefore, our assertion that the compressional component of the displacement field carries a well-defined elastic OAM is justified. The remaining contributions to the elastic OAM (from the transverse and hybrid components) also contain terms proportional to the azimuthal index $m$, but with additional factors (see supplemental material \cite{SM}) that leave them not fully quantised in the sense that they are only proportional to the azimuthal index. The physical significance of this for the exemplar case of guided waves in pipes is then that: (i) trivially, $\bm{M}_{L}^{o}\bm{\cdot}\hat{\bm{z}} = 0$ for both $L(0,n)$, $T(0,n)$ modes which is to be expected for axisymmetric modes; (ii) pure flexural modes with a constant angular profile are required to carry OAM. Conventional means of exciting these modes in pipes rely on either complex arrangements of transducers (e.g non-axisymmetric partial loading) or phased arrays \cite{shin1999guided,li2001excitation,li2002angular,rose2014ultrasonic,tang2017excitation}. Often many degenerate groups of flexural modes are excited simultaneously, including modes with both $\exp(\pm im\phi)$ components; the angular profile then changes with propagation distance due to modal superposition. As such there is zero average elastic OAM. Fortuitously, the recent advent of the elastic spiral phase pipe (eSPP), analogous to optical spiral phase plates \cite{chaplain2021elastic,beijersbergen1994helical}, enables arbitrary $F(m,n)$ modes to be efficiently excited, via mode conversion, which boast a constant angular profile along the pipe axis. 

We demonstrate in Fig.~\ref{fig:jzdisp}, via numerical calculation, that the elastic OAM associated with the dilatational potential for guided waves along a pipe carries a well-defined OAM. The associated orbital angular momentum flux density, at a constant plane in $z$, is given as $\bm{\mathcal{J}}_{L}\bm{\cdot}\hat{\bm{z}} = -\frac{\omega}{2}\mathfrak{Im}\int \boldsymbol{M}^{o}_{L}\bm{\cdot}\hat{\bm{z}} drd\theta$. For brevity we define $\mathcal{J}_{Lz} = \bm{\mathcal{J}}_{L}\bm{\cdot}\hat{\bm{z}}$. Figure~\ref{fig:jzdisp}(a) shows the dispersion curves for guided waves in an Aluminium pipe, evaluated using a spectral collocation method \cite{adamou2004spectral,quintanilla2015anisotropic,pavlakovic1997disperse} (corroborated with Finite Element computations \cite{marzani2008SAFE}, using the commercial software COMSOL Multiphysics \cite{comsol}\textsuperscript{\textregistered}), described in the supplemental material \cite{SM}; the eigensolutions give frequency as the eigenvalue, with the corresponding eigenvector containing the potential components $\left(\Phi,\Psi_r,\Psi_{\theta},\Psi_z\right)$. These are used to numerically evaluate the ratio of the compressional OAM flux density to the energy flux density of a compressional bulk wave, $\log|\frac{2\mathcal{J}_{Lz}}{\omega\rho c_p^2}|$, shown in Fig.~\ref{fig:jzdisp}(b) for the lowest branches of the first five flexural modes $F(m = 1 \dots 5,1)$. 

We now utilise an elastic spiral phase pipe (Fig.~\ref{fig:esp}) to show that elastic OAM can be transferred to a fluid in contact with the elastic material; shear waves are not supported in fluids and as such only the compressional motion of the elastic material couples strongly to the acoustic pressure field in the fluid. The OAM transfer is observed by the introduction of rotational motion within the fluid, exciting spiraling acoustic wave-fields.

\noindent \textit{OAM transfer.}|
The ability to transfer elastic OAM to a fluid is demonstrated numerically, via finite element simulations of an aluminium pipe partially submerged in water (Fig.~\ref{fig:Sim}. We excite a flexural $F(3,2)$ mode via mode conversion of a longitudinal $L(0,2)$ wave by passage through a suitably designed elastic spiral phase pipe (see supplemental material \cite{SM}). For a single frequency and wavevector, the compressional motion of the flexural mode within the pipe can be represented by a superposition of plane waves uniformly distributed over the circular aperture of the pipe. The plane wave components have mutual phases proportional to the azimuthal index $m$, endowed by the introduction of the eSPP. This compressional motion couples to the pressure field within the fluid at the submerged end of the pipe, producing rotating acoustic pressure fields. This is demonstrated in Figs.~\ref{fig:Sim}(b-c) that show a snapshot in time of the dilatational field, through the trace of the strain tensor $\tr(\varepsilon)$, at the submerged pipe end and the pressure field within the fluid. The dilatation is related to the compressional potential through $\Phi =  -(\nabla\bm{\cdot}\bm{u})/k_z^2 = -\tr(\varepsilon)/k_z^2$. We only consider a partially submerged pipe in order to neglect Franz-type waves \cite{frisk1975surface}, and note that the OAM transfer is viewed via the mechanical torque the compressional motion enacts on the fluid, not by the generation of acoustic Bessel beams known to carry OAM \cite{hefner1998acoustical,wunenburger2015acoustic,hong2015observation,bliokh2019spin,toftul2019acoustic}.

\noindent \textit{Conclusions.}|
We have shown that elastic waves with inclined phase fronts can carry an extrinsic orbital angular momentum; it has been proved that the compressional dilatational potential carries a well-defined contribution, proportional to the azimuthal index $m$. 
This result is reminiscent of the case of LG beams in optics where the electromagnetic wave equation, under the paraxial approximation, is satisfied by a complex scalar function describing the field distribution, proportional to the azimuthal mode index. It is this phase profile that gives rise, in both cases, to the well-defined OAM. 

The coupling of guided flexural waves in elastic pipes to acoustic pressure waves in fluids has been shown numerically through the compressional motion of the pipe. The implications are that the elastic OAM carried by the flexural modes can be transferred to acoustic pressure fields within a fluid. Inspired by the fact that optical LG laser modes have well-defined OAM, and that these modes are capable of being produced by spiral phase plates \cite{oemrawsingh2004production}, we leverage recent developments in elastic spiral phase pipes to generate the desired flexural pipe modes. These eSPP enables efficient mode conversion of longitudinal modes to arbitrary flexural modes, crucially of a single handedness, i.e that possess only one sign of $\exp(\pm i m \theta)$. In this way the compressional motion in the pipe acts as a continuous phased acoustic pressure source in the fluid, opposed to conventional discrete acoustic sources \cite{cromb2020amplification}. Harnessing the mechanical torques associated with the elastic OAM then promises to unlock applications across acoustic tweezers, non-destructive testing, ultrasonic motor design and acoustofluidic devices.

G.J.C gratefully acknowledges financial support from the Royal Commission for the Exhibition of 1851 in the form of a Research Fellowship. J.M.D.P \& R.V.C acknowledge the financial support from the  H2020 FET-proactive  project MetaVEH under grant agreement No. 952039.


\begin{thebibliography}{48}%
\makeatletter
\providecommand \@ifxundefined [1]{%
 \@ifx{#1\undefined}
}%
\providecommand \@ifnum [1]{%
 \ifnum #1\expandafter \@firstoftwo
 \else \expandafter \@secondoftwo
 \fi
}%
\providecommand \@ifx [1]{%
 \ifx #1\expandafter \@firstoftwo
 \else \expandafter \@secondoftwo
 \fi
}%
\providecommand \natexlab [1]{#1}%
\providecommand \enquote  [1]{``#1''}%
\providecommand \bibnamefont  [1]{#1}%
\providecommand \bibfnamefont [1]{#1}%
\providecommand \citenamefont [1]{#1}%
\providecommand \href@noop [0]{\@secondoftwo}%
\providecommand \href [0]{\begingroup \@sanitize@url \@href}%
\providecommand \@href[1]{\@@startlink{#1}\@@href}%
\providecommand \@@href[1]{\endgroup#1\@@endlink}%
\providecommand \@sanitize@url [0]{\catcode `\\12\catcode `\$12\catcode
  `\&12\catcode `\#12\catcode `\^12\catcode `\_12\catcode `\%12\relax}%
\providecommand \@@startlink[1]{}%
\providecommand \@@endlink[0]{}%
\providecommand \url  [0]{\begingroup\@sanitize@url \@url }%
\providecommand \@url [1]{\endgroup\@href {#1}{\urlprefix }}%
\providecommand \urlprefix  [0]{URL }%
\providecommand \Eprint [0]{\href }%
\providecommand \doibase [0]{https://doi.org/}%
\providecommand \selectlanguage [0]{\@gobble}%
\providecommand \bibinfo  [0]{\@secondoftwo}%
\providecommand \bibfield  [0]{\@secondoftwo}%
\providecommand \translation [1]{[#1]}%
\providecommand \BibitemOpen [0]{}%
\providecommand \bibitemStop [0]{}%
\providecommand \bibitemNoStop [0]{.\EOS\space}%
\providecommand \EOS [0]{\spacefactor3000\relax}%
\providecommand \BibitemShut  [1]{\csname bibitem#1\endcsname}%
\let\auto@bib@innerbib\@empty
\bibitem [{\citenamefont {Allen}\ \emph {et~al.}(1992)\citenamefont {Allen},
  \citenamefont {Beijersbergen}, \citenamefont {Spreeuw},\ and\ \citenamefont
  {Woerdman}}]{allen1992orbital}%
  \BibitemOpen
  \bibfield  {author} {\bibinfo {author} {\bibfnamefont {L.}~\bibnamefont
  {Allen}}, \bibinfo {author} {\bibfnamefont {M.~W.}\ \bibnamefont
  {Beijersbergen}}, \bibinfo {author} {\bibfnamefont {R.~J.~C.}~\bibnamefont
  {Spreeuw}},\ and\ \bibinfo {author} {\bibfnamefont {J.~P.}~\bibnamefont
  {Woerdman}},\ }\bibfield  {title} {\bibinfo {title} {Orbital angular momentum
  of light and the transformation of {L}aguerre-{G}aussian laser modes},\
  }\href@noop {} {\bibfield  {journal} {\bibinfo  {journal} {Physical review
  A}\ }\textbf {\bibinfo {volume} {45}},\ \bibinfo {pages} {8185} (\bibinfo
  {year} {1992})}\BibitemShut {NoStop}%
\bibitem [{\citenamefont {Jackson}(1962)}]{jackson1999classical}%
  \BibitemOpen
  \bibfield  {author} {\bibinfo {author} {\bibfnamefont {J.~D.}\ \bibnamefont
  {Jackson}},\ }\href@noop {} {\emph {\bibinfo {title} {Classical
  electrodynamics}}}\ (\bibinfo  {publisher} {Wiley, New York},\ \bibinfo
  {year} {1962})\BibitemShut {NoStop}%
\bibitem [{\citenamefont {He}\ \emph {et~al.}(1995)\citenamefont {He},
  \citenamefont {Friese}, \citenamefont {Heckenberg},\ and\ \citenamefont
  {Rubinsztein-Dunlop}}]{he1995direct}%
  \BibitemOpen
  \bibfield  {author} {\bibinfo {author} {\bibfnamefont {H.}~\bibnamefont
  {He}}, \bibinfo {author} {\bibfnamefont {M.~E.~J.}~\bibnamefont {Friese}}, \bibinfo
  {author} {\bibfnamefont {N.~R.}~\bibnamefont {Heckenberg}},\ and\ \bibinfo
  {author} {\bibfnamefont {H.}~\bibnamefont {Rubinsztein-Dunlop}},\ }\bibfield
  {title} {\bibinfo {title} {Direct observation of transfer of angular momentum
  to absorptive particles from a laser beam with a phase singularity},\
  }\href@noop {} {\bibfield  {journal} {\bibinfo  {journal} {Physical review
  letters}\ }\textbf {\bibinfo {volume} {75}},\ \bibinfo {pages} {826}
  (\bibinfo {year} {1995})}\BibitemShut {NoStop}%
\bibitem [{\citenamefont {O'Neil}\ \emph {et~al.}(2002)\citenamefont {O'Neil},
  \citenamefont {MacVicar}, \citenamefont {Allen},\ and\ \citenamefont
  {Padgett}}]{o2002intrinsic}%
  \BibitemOpen
  \bibfield  {author} {\bibinfo {author} {\bibfnamefont {A.~T.}~\bibnamefont
  {O'Neil}}, \bibinfo {author} {\bibfnamefont {I.}~\bibnamefont {MacVicar}},
  \bibinfo {author} {\bibfnamefont {L.}~\bibnamefont {Allen}},\ and\ \bibinfo
  {author} {\bibfnamefont {M.~J.}~\bibnamefont {Padgett}},\ }\bibfield  {title}
  {\bibinfo {title} {Intrinsic and extrinsic nature of the orbital angular
  momentum of a light beam},\ }\href@noop {} {\bibfield  {journal} {\bibinfo
  {journal} {Physical review letters}\ }\textbf {\bibinfo {volume} {88}},\
  \bibinfo {pages} {053601} (\bibinfo {year} {2002})}\BibitemShut {NoStop}%
\bibitem [{\citenamefont {Beth}(1936)}]{beth1936mechanical}%
  \BibitemOpen
  \bibfield  {author} {\bibinfo {author} {\bibfnamefont {R.~A.}\ \bibnamefont
  {Beth}},\ }\bibfield  {title} {\bibinfo {title} {Mechanical detection and
  measurement of the angular momentum of light},\ }\href@noop {} {\bibfield
  {journal} {\bibinfo  {journal} {Physical Review}\ }\textbf {\bibinfo {volume}
  {50}},\ \bibinfo {pages} {115} (\bibinfo {year} {1936})}\BibitemShut
  {NoStop}%
\bibitem [{\citenamefont {Poynting}(1909)}]{poynting1909wave}%
  \BibitemOpen
  \bibfield  {author} {\bibinfo {author} {\bibfnamefont {J.~H.}\ \bibnamefont
  {Poynting}},\ }\bibfield  {title} {\bibinfo {title} {The wave motion of a
  revolving shaft, and a suggestion as to the angular momentum in a beam of
  circularly polarised light},\ }\href@noop {} {\bibfield  {journal} {\bibinfo
  {journal} {Proceedings of the Royal Society of London. Series A, Containing
  Papers of a Mathematical and Physical Character}\ }\textbf {\bibinfo {volume}
  {82}},\ \bibinfo {pages} {560} (\bibinfo {year} {1909})}\BibitemShut
  {NoStop}%
\bibitem [{\citenamefont {Barnett}(2001)}]{barnett2001optical}%
  \BibitemOpen
  \bibfield  {author} {\bibinfo {author} {\bibfnamefont {S.~M.}\ \bibnamefont
  {Barnett}},\ }\bibfield  {title} {\bibinfo {title} {Optical angular-momentum
  flux},\ }\href@noop {} {\bibfield  {journal} {\bibinfo  {journal} {Journal of
  Optics B: Quantum and Semiclassical Optics}\ }\textbf {\bibinfo {volume}
  {4}},\ \bibinfo {pages} {S7} (\bibinfo {year} {2001})}\BibitemShut {NoStop}%
\bibitem [{\citenamefont {Barnett}(2010)}]{barnett2010rotation}%
  \BibitemOpen
  \bibfield  {author} {\bibinfo {author} {\bibfnamefont {S.~M.}\ \bibnamefont
  {Barnett}},\ }\bibfield  {title} {\bibinfo {title} {Rotation of
  electromagnetic fields and the nature of optical angular momentum},\
  }\href@noop {} {\bibfield  {journal} {\bibinfo  {journal} {Journal of modern
  optics}\ }\textbf {\bibinfo {volume} {57}},\ \bibinfo {pages} {1339}
  (\bibinfo {year} {2010})}\BibitemShut {NoStop}%
\bibitem [{\citenamefont {Allen}\ \emph {et~al.}(2016)\citenamefont {Allen},
  \citenamefont {Barnett},\ and\ \citenamefont {Padgett}}]{allen2016optical}%
  \BibitemOpen
  \bibfield  {author} {\bibinfo {author} {\bibfnamefont {L.}~\bibnamefont
  {Allen}}, \bibinfo {author} {\bibfnamefont {S.~M.}\ \bibnamefont {Barnett}},\
  and\ \bibinfo {author} {\bibfnamefont {M.~J.}\ \bibnamefont {Padgett}},\
  }\href@noop {} {\emph {\bibinfo {title} {Optical angular momentum}}}\
  (\bibinfo  {publisher} {CRC press},\ \bibinfo {year} {2016})\BibitemShut
  {NoStop}%
\bibitem [{\citenamefont {Garc{\'e}s-Ch{\'a}vez}\ \emph
  {et~al.}(2003)\citenamefont {Garc{\'e}s-Ch{\'a}vez}, \citenamefont {McGloin},
  \citenamefont {Padgett}, \citenamefont {Dultz}, \citenamefont {Schmitzer},\
  and\ \citenamefont {Dholakia}}]{garces2003observation}%
  \BibitemOpen
  \bibfield  {author} {\bibinfo {author} {\bibfnamefont {V.}~\bibnamefont
  {Garc{\'e}s-Ch{\'a}vez}}, \bibinfo {author} {\bibfnamefont {D.}~\bibnamefont
  {McGloin}}, \bibinfo {author} {\bibfnamefont {M.~J.}~\bibnamefont {Padgett}},
  \bibinfo {author} {\bibfnamefont {W.}~\bibnamefont {Dultz}}, \bibinfo
  {author} {\bibfnamefont {H.}~\bibnamefont {Schmitzer}},\ and\ \bibinfo
  {author} {\bibfnamefont {K.}~\bibnamefont {Dholakia}},\ }\bibfield  {title}
  {\bibinfo {title} {Observation of the transfer of the local angular momentum
  density of a multiringed light beam to an optically trapped particle},\
  }\href@noop {} {\bibfield  {journal} {\bibinfo  {journal} {Physical review
  letters}\ }\textbf {\bibinfo {volume} {91}},\ \bibinfo {pages} {093602}
  (\bibinfo {year} {2003})}\BibitemShut {NoStop}%
\bibitem [{\citenamefont {Yao}\ and\ \citenamefont
  {Padgett}(2011)}]{yao2011orbital}%
  \BibitemOpen
  \bibfield  {author} {\bibinfo {author} {\bibfnamefont {A.~M.}\ \bibnamefont
  {Yao}}\ and\ \bibinfo {author} {\bibfnamefont {M.~J.}\ \bibnamefont
  {Padgett}},\ }\bibfield  {title} {\bibinfo {title} {Orbital angular momentum:
  origins, behavior and applications},\ }\href@noop {} {\bibfield  {journal}
  {\bibinfo  {journal} {Advances in Optics and Photonics}\ }\textbf {\bibinfo
  {volume} {3}},\ \bibinfo {pages} {161} (\bibinfo {year} {2011})}\BibitemShut
  {NoStop}%
\bibitem [{\citenamefont {Willner}\ \emph {et~al.}(2015)\citenamefont
  {Willner}, \citenamefont {Huang}, \citenamefont {Yan}, \citenamefont {Ren},
  \citenamefont {Ahmed}, \citenamefont {Xie}, \citenamefont {Bao},
  \citenamefont {Li}, \citenamefont {Cao}, \citenamefont {Zhao} \emph
  {et~al.}}]{willner2015optical}%
  \BibitemOpen
  \bibfield  {author} {\bibinfo {author} {\bibfnamefont {A.~E.}\ \bibnamefont
  {Willner}}, \bibinfo {author} {\bibfnamefont {H.}~\bibnamefont {Huang}},
  \bibinfo {author} {\bibfnamefont {Y.}~\bibnamefont {Yan}}, \bibinfo {author}
  {\bibfnamefont {Y.}~\bibnamefont {Ren}}, \bibinfo {author} {\bibfnamefont
  {N.}~\bibnamefont {Ahmed}}, \bibinfo {author} {\bibfnamefont
  {G.}~\bibnamefont {Xie}}, \bibinfo {author} {\bibfnamefont {C.}~\bibnamefont
  {Bao}}, \bibinfo {author} {\bibfnamefont {L.}~\bibnamefont {Li}}, \bibinfo
  {author} {\bibfnamefont {Y.}~\bibnamefont {Cao}}, \bibinfo {author}
  {\bibfnamefont {Z.}~\bibnamefont {Zhao}}, \emph {et~al.},\ }\bibfield
  {title} {\bibinfo {title} {Optical communications using orbital angular
  momentum beams},\ }\href@noop {} {\bibfield  {journal} {\bibinfo  {journal}
  {Advances in Optics and Photonics}\ }\textbf {\bibinfo {volume} {7}},\
  \bibinfo {pages} {66} (\bibinfo {year} {2015})}\BibitemShut {NoStop}%
\bibitem [{\citenamefont {Bliokh}\ and\ \citenamefont
  {Nori}(2015)}]{bliokh2015transverse}%
  \BibitemOpen
  \bibfield  {author} {\bibinfo {author} {\bibfnamefont {K.~Y.}\ \bibnamefont
  {Bliokh}}\ and\ \bibinfo {author} {\bibfnamefont {F.}~\bibnamefont {Nori}},\
  }\bibfield  {title} {\bibinfo {title} {Transverse and longitudinal angular
  momenta of light},\ }\href@noop {} {\bibfield  {journal} {\bibinfo  {journal}
  {Physics Reports}\ }\textbf {\bibinfo {volume} {592}},\ \bibinfo {pages} {1}
  (\bibinfo {year} {2015})}\BibitemShut {NoStop}%
\bibitem [{\citenamefont {Barnett}\ \emph {et~al.}(2017)\citenamefont
  {Barnett}, \citenamefont {Babiker},\ and\ \citenamefont
  {Padgett}}]{barnett2017optical}%
  \BibitemOpen
  \bibfield  {author} {\bibinfo {author} {\bibfnamefont {S.~M.}\ \bibnamefont
  {Barnett}}, \bibinfo {author} {\bibfnamefont {M.}~\bibnamefont {Babiker}},\
  and\ \bibinfo {author} {\bibfnamefont {M.~J.}\ \bibnamefont {Padgett}},\
  }\bibfield  {title} {\bibinfo {title} {Optical orbital angular momentum},\
  }\href@noop {} {\bibfield  {journal} {\bibinfo  {journal} {Phil. Trans. R.
  Soc. A.}\ }\textbf {\bibinfo {volume} {375}},\ \bibinfo {pages} {20150444}
  (\bibinfo {year} {2017})}\BibitemShut {NoStop}%
\bibitem [{\citenamefont {Padgett}(2017)}]{padgett2017orbital}%
  \BibitemOpen
  \bibfield  {author} {\bibinfo {author} {\bibfnamefont {M.~J.}\ \bibnamefont
  {Padgett}},\ }\bibfield  {title} {\bibinfo {title} {Orbital angular momentum
  25 years on},\ }\href@noop {} {\bibfield  {journal} {\bibinfo  {journal}
  {Optics express}\ }\textbf {\bibinfo {volume} {25}},\ \bibinfo {pages}
  {11265} (\bibinfo {year} {2017})}\BibitemShut {NoStop}%
\bibitem [{\citenamefont {Chen}\ \emph {et~al.}(2019)\citenamefont {Chen},
  \citenamefont {Zhou}, \citenamefont {Moretti}, \citenamefont {Wang},\ and\
  \citenamefont {Li}}]{chen2019orbital}%
  \BibitemOpen
  \bibfield  {author} {\bibinfo {author} {\bibfnamefont {R.}~\bibnamefont
  {Chen}}, \bibinfo {author} {\bibfnamefont {H.}~\bibnamefont {Zhou}}, \bibinfo
  {author} {\bibfnamefont {M.}~\bibnamefont {Moretti}}, \bibinfo {author}
  {\bibfnamefont {X.}~\bibnamefont {Wang}},\ and\ \bibinfo {author}
  {\bibfnamefont {J.}~\bibnamefont {Li}},\ }\bibfield  {title} {\bibinfo
  {title} {Orbital angular momentum waves: generation, detection, and emerging
  applications},\ }\href@noop {} {\bibfield  {journal} {\bibinfo  {journal}
  {IEEE Communications Surveys \& Tutorials}\ }\textbf {\bibinfo {volume}
  {22}},\ \bibinfo {pages} {840} (\bibinfo {year} {2019})}\BibitemShut
  {NoStop}%
\bibitem [{\citenamefont {Long}\ \emph {et~al.}(2018)\citenamefont {Long},
  \citenamefont {Ren},\ and\ \citenamefont {Chen}}]{long2018intrinsic}%
  \BibitemOpen
  \bibfield  {author} {\bibinfo {author} {\bibfnamefont {Y.}~\bibnamefont
  {Long}}, \bibinfo {author} {\bibfnamefont {J.}~\bibnamefont {Ren}},\ and\
  \bibinfo {author} {\bibfnamefont {H.}~\bibnamefont {Chen}},\ }\bibfield
  {title} {\bibinfo {title} {Intrinsic spin of elastic waves},\ }\href@noop {}
  {\bibfield  {journal} {\bibinfo  {journal} {Proceedings of the National
  Academy of Sciences}\ }\textbf {\bibinfo {volume} {115}},\ \bibinfo {pages}
  {9951} (\bibinfo {year} {2018})}\BibitemShut {NoStop}%
\bibitem [{\citenamefont {Deymier}\ \emph {et~al.}(2018)\citenamefont
  {Deymier}, \citenamefont {Runge}, \citenamefont {Vasseur}, \citenamefont
  {Hladky},\ and\ \citenamefont {Lucas}}]{deymier2018elastic}%
  \BibitemOpen
  \bibfield  {author} {\bibinfo {author} {\bibfnamefont {P.}~\bibnamefont
  {Deymier}}, \bibinfo {author} {\bibfnamefont {K.}~\bibnamefont {Runge}},
  \bibinfo {author} {\bibfnamefont {J.}~\bibnamefont {Vasseur}}, \bibinfo
  {author} {\bibfnamefont {A.}~\bibnamefont {Hladky}},\ and\ \bibinfo {author}
  {\bibfnamefont {P.}~\bibnamefont {Lucas}},\ }\bibfield  {title} {\bibinfo
  {title} {Elastic waves with correlated directional and orbital angular
  momentum degrees of freedom},\ }\href@noop {} {\bibfield  {journal} {\bibinfo
   {journal} {Journal of Physics B: Atomic, Molecular and Optical Physics}\
  }\textbf {\bibinfo {volume} {51}},\ \bibinfo {pages} {135301} (\bibinfo
  {year} {2018})}\BibitemShut {NoStop}%
\bibitem [{\citenamefont {Eshelby}(1951)}]{eshelby1951force}%
  \BibitemOpen
  \bibfield  {author} {\bibinfo {author} {\bibfnamefont {J.~D.}\ \bibnamefont
  {Eshelby}},\ }\bibfield  {title} {\bibinfo {title} {The force on an elastic
  singularity},\ }\href@noop {} {\bibfield  {journal} {\bibinfo  {journal}
  {Philosophical Transactions of the Royal Society of London. Series A,
  Mathematical and Physical Sciences}\ }\textbf {\bibinfo {volume} {244}},\
  \bibinfo {pages} {87} (\bibinfo {year} {1951})}\BibitemShut {NoStop}%
\bibitem [{\citenamefont {Eshelby}(1975)}]{eshelby1975elastic}%
  \BibitemOpen
  \bibfield  {author} {\bibinfo {author} {\bibfnamefont {J.}~\bibnamefont
  {Eshelby}},\ }\bibfield  {title} {\bibinfo {title} {The elastic
  energy-momentum tensor},\ }\href@noop {} {\bibfield  {journal} {\bibinfo
  {journal} {Journal of elasticity}\ }\textbf {\bibinfo {volume} {5}},\
  \bibinfo {pages} {321} (\bibinfo {year} {1975})}\BibitemShut {NoStop}%
\bibitem [{\citenamefont {Thielheim}(1967)}]{thielheim1967note}%
  \BibitemOpen
  \bibfield  {author} {\bibinfo {author} {\bibfnamefont {K.~O.}\ \bibnamefont
  {Thielheim}},\ }\bibfield  {title} {\bibinfo {title} {Note on classical
  fields of higher order},\ }\href@noop {} {\bibfield  {journal} {\bibinfo
  {journal} {Proceedings of the Physical Society (1958-1967)}\ }\textbf
  {\bibinfo {volume} {91}},\ \bibinfo {pages} {798} (\bibinfo {year}
  {1967})}\BibitemShut {NoStop}%
\bibitem [{\citenamefont {Lazar}\ and\ \citenamefont
  {Kirchner}(2007)}]{lazar2007eshelby}%
  \BibitemOpen
  \bibfield  {author} {\bibinfo {author} {\bibfnamefont {M.}~\bibnamefont
  {Lazar}}\ and\ \bibinfo {author} {\bibfnamefont {H.~O.}\ \bibnamefont
  {Kirchner}},\ }\bibfield  {title} {\bibinfo {title} {The {E}shelby stress
  tensor, angular momentum tensor and dilatation flux in gradient elasticity},\
  }\href@noop {} {\bibfield  {journal} {\bibinfo  {journal} {International
  Journal of Solids and Structures}\ }\textbf {\bibinfo {volume} {44}},\
  \bibinfo {pages} {2477} (\bibinfo {year} {2007})}\BibitemShut {NoStop}%
\bibitem [{\citenamefont {Landau}\ and\ \citenamefont
  {Lifshitz}(1959)}]{landau1959course}%
  \BibitemOpen
  \bibfield  {author} {\bibinfo {author} {\bibfnamefont {L.~D.}\ \bibnamefont
  {Landau}}\ and\ \bibinfo {author} {\bibfnamefont {E.~M.}\ \bibnamefont
  {Lifshitz}},\ }\href@noop {} {\emph {\bibinfo {title} {Course of Theoretical
  Physics Vol 7: Theory and Elasticity}}}\ (\bibinfo  {publisher} {Pergamon
  press},\ \bibinfo {year} {1959})\BibitemShut {NoStop}%
\bibitem [{\citenamefont {Noether}(1971)}]{noether1971invariant}%
  \BibitemOpen
  \bibfield  {author} {\bibinfo {author} {\bibfnamefont {E.}~\bibnamefont
  {Noether}},\ }\bibfield  {title} {\bibinfo {title} {Invariant variation
  problems},\ }\href@noop {} {\bibfield  {journal} {\bibinfo  {journal}
  {Transport theory and statistical physics}\ }\textbf {\bibinfo {volume}
  {1}},\ \bibinfo {pages} {186} (\bibinfo {year} {1971})}\BibitemShut {NoStop}%
\bibitem [{SM()}]{SM}%
  \BibitemOpen
  \bibfield  {title} {\bibinfo {title} {See supplemental material at xxxx for
  more details on the theory of laguerre-gaussian beams, analytical and
  numerical solutions for guided waves in pipes},\ }\href@noop {} {\
  }\BibitemShut {NoStop}%
\bibitem [{\citenamefont {Gazis}(1959{\natexlab{a}})}]{gazis1959a}%
  \BibitemOpen
  \bibfield  {author} {\bibinfo {author} {\bibfnamefont {D.~C.}\ \bibnamefont
  {Gazis}},\ }\bibfield  {title} {\bibinfo {title} {Three-dimensional
  investigation of the propagation of waves in hollow circular cylinders. i.
  analytical foundation},\ }\href@noop {} {\bibfield  {journal} {\bibinfo
  {journal} {The Journal of the Acoustical Society of America}\ }\textbf
  {\bibinfo {volume} {31}},\ \bibinfo {pages} {568} (\bibinfo {year}
  {1959}{\natexlab{a}})}\BibitemShut {NoStop}%
\bibitem [{\citenamefont {Gazis}(1959{\natexlab{b}})}]{gazis1959b}%
  \BibitemOpen
  \bibfield  {author} {\bibinfo {author} {\bibfnamefont {D.~C.}\ \bibnamefont
  {Gazis}},\ }\bibfield  {title} {\bibinfo {title} {Three-dimensional
  investigation of the propagation of waves in hollow circular cylinders. ii.
  numerical results},\ }\href@noop {} {\bibfield  {journal} {\bibinfo
  {journal} {The Journal of the Acoustical Society of America}\ }\textbf
  {\bibinfo {volume} {31}},\ \bibinfo {pages} {573} (\bibinfo {year}
  {1959}{\natexlab{b}})}\BibitemShut {NoStop}%
\bibitem [{\citenamefont {Silk}\ and\ \citenamefont
  {Bainton}(1979)}]{silk1979propagation}%
  \BibitemOpen
  \bibfield  {author} {\bibinfo {author} {\bibfnamefont {M.}~\bibnamefont
  {Silk}}\ and\ \bibinfo {author} {\bibfnamefont {K.}~\bibnamefont {Bainton}},\
  }\bibfield  {title} {\bibinfo {title} {The propagation in metal tubing of
  ultrasonic wave modes equivalent to {L}amb waves},\ }\href@noop {} {\bibfield
   {journal} {\bibinfo  {journal} {Ultrasonics}\ }\textbf {\bibinfo {volume}
  {17}},\ \bibinfo {pages} {11} (\bibinfo {year} {1979})}\BibitemShut {NoStop}%
\bibitem [{\citenamefont {Shin}\ and\ \citenamefont
  {Rose}(1999)}]{shin1999guided}%
  \BibitemOpen
  \bibfield  {author} {\bibinfo {author} {\bibfnamefont {H.~J.}\ \bibnamefont
  {Shin}}\ and\ \bibinfo {author} {\bibfnamefont {J.~L.}\ \bibnamefont
  {Rose}},\ }\bibfield  {title} {\bibinfo {title} {Guided waves by axisymmetric
  and non-axisymmetric surface loading on hollow cylinders},\ }\href@noop {}
  {\bibfield  {journal} {\bibinfo  {journal} {Ultrasonics}\ }\textbf {\bibinfo
  {volume} {37}},\ \bibinfo {pages} {355} (\bibinfo {year} {1999})}\BibitemShut
  {NoStop}%
\bibitem [{\citenamefont {Li}\ and\ \citenamefont
  {Rose}(2001)}]{li2001excitation}%
  \BibitemOpen
  \bibfield  {author} {\bibinfo {author} {\bibfnamefont {J.}~\bibnamefont
  {Li}}\ and\ \bibinfo {author} {\bibfnamefont {J.~L.}\ \bibnamefont {Rose}},\
  }\bibfield  {title} {\bibinfo {title} {Excitation and propagation of
  non-axisymmetric guided waves in a hollow cylinder},\ }\href@noop {}
  {\bibfield  {journal} {\bibinfo  {journal} {The Journal of the Acoustical
  Society of America}\ }\textbf {\bibinfo {volume} {109}},\ \bibinfo {pages}
  {457} (\bibinfo {year} {2001})}\BibitemShut {NoStop}%
\bibitem [{\citenamefont {Li}\ and\ \citenamefont
  {Rose}(2002)}]{li2002angular}%
  \BibitemOpen
  \bibfield  {author} {\bibinfo {author} {\bibfnamefont {J.}~\bibnamefont
  {Li}}\ and\ \bibinfo {author} {\bibfnamefont {J.~L.}\ \bibnamefont {Rose}},\
  }\bibfield  {title} {\bibinfo {title} {Angular-profile tuning of guided waves
  in hollow cylinders using a circumferential phased array},\ }\href@noop {}
  {\bibfield  {journal} {\bibinfo  {journal} {IEEE transactions on ultrasonics,
  ferroelectrics, and frequency control}\ }\textbf {\bibinfo {volume} {49}},\
  \bibinfo {pages} {1720} (\bibinfo {year} {2002})}\BibitemShut {NoStop}%
\bibitem [{\citenamefont {Rose}(2014)}]{rose2014ultrasonic}%
  \BibitemOpen
  \bibfield  {author} {\bibinfo {author} {\bibfnamefont {J.~L.}\ \bibnamefont
  {Rose}},\ }\href@noop {} {\emph {\bibinfo {title} {Ultrasonic guided waves in
  solid media}}}\ (\bibinfo  {publisher} {Cambridge University Press},\
  \bibinfo {year} {2014})\BibitemShut {NoStop}%
\bibitem [{\citenamefont {Tang}\ and\ \citenamefont
  {Wu}(2017)}]{tang2017excitation}%
  \BibitemOpen
  \bibfield  {author} {\bibinfo {author} {\bibfnamefont {L.}~\bibnamefont
  {Tang}}\ and\ \bibinfo {author} {\bibfnamefont {B.}~\bibnamefont {Wu}},\
  }\bibfield  {title} {\bibinfo {title} {Excitation mechanism of
  flexural-guided wave modes {F}(1, 2) and {F}(1, 3) in pipes},\ }\href@noop {}
  {\bibfield  {journal} {\bibinfo  {journal} {Journal of Nondestructive
  Evaluation}\ }\textbf {\bibinfo {volume} {36}},\ \bibinfo {pages} {1}
  (\bibinfo {year} {2017})}\BibitemShut {NoStop}%
\bibitem [{\citenamefont {Chaplain}\ and\ \citenamefont
  {Ponti}(2021)}]{chaplain2021elastic}%
  \BibitemOpen
  \bibfield  {author} {\bibinfo {author} {\bibfnamefont {G.~J.}\ \bibnamefont
  {Chaplain}}\ and\ \bibinfo {author} {\bibfnamefont {J.~M.}\ \bibnamefont
  {De Ponti}},\ }\href@noop {} {\bibinfo {title} {The elastic spiral phase pipe},\ }\href@noop {}
  {\bibfield  {journal} {\bibinfo  {journal} {Journal of Sound and Vibration}\ }\textbf {\bibinfo {volume} {523}},\ \bibinfo {pages} {116718}
  (\bibinfo {year} {2022})}\BibitemShut {NoStop}%
\bibitem [{\citenamefont {Beijersbergen}\ \emph {et~al.}(1994)\citenamefont
  {Beijersbergen}, \citenamefont {Coerwinkel}, \citenamefont {Kristensen},\
  and\ \citenamefont {Woerdman}}]{beijersbergen1994helical}%
  \BibitemOpen
  \bibfield  {author} {\bibinfo {author} {\bibfnamefont {M.}~\bibnamefont
  {Beijersbergen}}, \bibinfo {author} {\bibfnamefont {R.}~\bibnamefont
  {Coerwinkel}}, \bibinfo {author} {\bibfnamefont {M.}~\bibnamefont
  {Kristensen}},\ and\ \bibinfo {author} {\bibfnamefont {J.}~\bibnamefont
  {Woerdman}},\ }\bibfield  {title} {\bibinfo {title} {Helical-wavefront laser
  beams produced with a spiral phaseplate},\ }\href@noop {} {\bibfield
  {journal} {\bibinfo  {journal} {Optics communications}\ }\textbf {\bibinfo
  {volume} {112}},\ \bibinfo {pages} {321} (\bibinfo {year}
  {1994})}\BibitemShut {NoStop}%
\bibitem [{\citenamefont {Adamou}\ and\ \citenamefont
  {Craster}(2004)}]{adamou2004spectral}%
  \BibitemOpen
  \bibfield  {author} {\bibinfo {author} {\bibfnamefont {A.}~\bibnamefont
  {Adamou}}\ and\ \bibinfo {author} {\bibfnamefont {R.}~\bibnamefont
  {Craster}},\ }\bibfield  {title} {\bibinfo {title} {Spectral methods for
  modelling guided waves in elastic media},\ }\href@noop {} {\bibfield
  {journal} {\bibinfo  {journal} {The Journal of the Acoustical Society of
  America}\ }\textbf {\bibinfo {volume} {116}},\ \bibinfo {pages} {1524}
  (\bibinfo {year} {2004})}\BibitemShut {NoStop}%
\bibitem [{\citenamefont {Quintanilla}\ \emph {et~al.}(2015)\citenamefont
  {Quintanilla}, \citenamefont {Lowe},\ and\ \citenamefont
  {Craster}}]{quintanilla2015anisotropic}%
  \BibitemOpen
  \bibfield  {author} {\bibinfo {author} {\bibfnamefont {F.~H.}\ \bibnamefont
  {Quintanilla}}, \bibinfo {author} {\bibfnamefont {M.~J.~S.}\ \bibnamefont
  {Lowe}},\ and\ \bibinfo {author} {\bibfnamefont {R.~V.}\ \bibnamefont
  {Craster}},\ }\bibfield  {title} {\bibinfo {title} {Modeling guided elastic
  waves in generally anisotropic media using a spectral collocation method},\
  }\href {https://doi.org/10.1121/1.4913777} {\bibfield  {journal} {\bibinfo
  {journal} {The Journal of the Acoustical Society of America}\ }\textbf
  {\bibinfo {volume} {137}},\ \bibinfo {pages} {1180} (\bibinfo {year}
  {2015})},\ \Eprint {https://arxiv.org/abs/https://doi.org/10.1121/1.4913777}
  {https://doi.org/10.1121/1.4913777} \BibitemShut {NoStop}%
\bibitem [{\citenamefont {Pavlakovic}\ \emph {et~al.}(1997)\citenamefont
  {Pavlakovic}, \citenamefont {Lowe}, \citenamefont {Alleyne},\ and\
  \citenamefont {Cawley}}]{pavlakovic1997disperse}%
  \BibitemOpen
  \bibfield  {author} {\bibinfo {author} {\bibfnamefont {B.}~\bibnamefont
  {Pavlakovic}}, \bibinfo {author} {\bibfnamefont {M.}~\bibnamefont {Lowe}},
  \bibinfo {author} {\bibfnamefont {D.}~\bibnamefont {Alleyne}},\ and\ \bibinfo
  {author} {\bibfnamefont {P.}~\bibnamefont {Cawley}},\ }\bibfield  {title}
  {\bibinfo {title} {Disperse: A general purpose program for creating
  dispersion curves},\ }in\ \href@noop {} {\emph {\bibinfo {booktitle} {Review
  of progress in quantitative nondestructive evaluation}}}\ (\bibinfo
  {publisher} {Springer},\ \bibinfo {year} {1997})\ pp.\ \bibinfo {pages}
  {185--192}\BibitemShut {NoStop}%
\bibitem [{\citenamefont {Marzani}\ \emph {et~al.}(2008)\citenamefont
  {Marzani}, \citenamefont {Viola}, \citenamefont {Bartoli}, \citenamefont
  {{Lanza di Scalea}},\ and\ \citenamefont {Rizzo}}]{marzani2008SAFE}%
  \BibitemOpen
  \bibfield  {author} {\bibinfo {author} {\bibfnamefont {A.}~\bibnamefont
  {Marzani}}, \bibinfo {author} {\bibfnamefont {E.}~\bibnamefont {Viola}},
  \bibinfo {author} {\bibfnamefont {I.}~\bibnamefont {Bartoli}}, \bibinfo
  {author} {\bibfnamefont {F.}~\bibnamefont {{Lanza di Scalea}}},\ and\
  \bibinfo {author} {\bibfnamefont {P.}~\bibnamefont {Rizzo}},\ }\bibfield
  {title} {\bibinfo {title} {A semi-analytical finite element formulation for
  modeling stress wave propagation in axisymmetric damped waveguides},\ }\href
  {https://doi.org/https://doi.org/10.1016/j.jsv.2008.04.028} {\bibfield
  {journal} {\bibinfo  {journal} {Journal of Sound and Vibration}\ }\textbf
  {\bibinfo {volume} {318}},\ \bibinfo {pages} {488} (\bibinfo {year}
  {2008})}\BibitemShut {NoStop}%
\bibitem [{com(2021)}]{comsol}%
  \BibitemOpen
  \href {http://www.comsol.com/products/multiphysics/} {\bibinfo {title}
  {\uppercase{COMSOL M}ultiphysics\textsuperscript{\textregistered} reference
  manual, version 5.6, www.comsol.com, \uppercase{COMSOL AB},
  \uppercase{S}tockholm, \uppercase{S}weden}} (\bibinfo {year}
  {2021})\BibitemShut {NoStop}%
\bibitem [{\citenamefont {Frisk}\ \emph {et~al.}(1975)\citenamefont {Frisk},
  \citenamefont {Dickey},\ and\ \citenamefont
  {{\"U}berall}}]{frisk1975surface}%
  \BibitemOpen
  \bibfield  {author} {\bibinfo {author} {\bibfnamefont {G.}~\bibnamefont
  {Frisk}}, \bibinfo {author} {\bibfnamefont {J.}~\bibnamefont {Dickey}},\ and\
  \bibinfo {author} {\bibfnamefont {H.}~\bibnamefont {{\"U}berall}},\
  }\bibfield  {title} {\bibinfo {title} {Surface wave modes on elastic
  cylinders},\ }\href@noop {} {\bibfield  {journal} {\bibinfo  {journal} {The
  Journal of the Acoustical Society of America}\ }\textbf {\bibinfo {volume}
  {58}},\ \bibinfo {pages} {996} (\bibinfo {year} {1975})}\BibitemShut
  {NoStop}%
\bibitem [{\citenamefont {Hefner}\ and\ \citenamefont
  {Marston}(1998)}]{hefner1998acoustical}%
  \BibitemOpen
  \bibfield  {author} {\bibinfo {author} {\bibfnamefont {B.~T.}\ \bibnamefont
  {Hefner}}\ and\ \bibinfo {author} {\bibfnamefont {P.~L.}\ \bibnamefont
  {Marston}},\ }\bibfield  {title} {\bibinfo {title} {Acoustical helicoidal
  waves and laguerre-gaussian beams: Applications to scattering and to angular
  momentum transport},\ }\href@noop {} {\bibfield  {journal} {\bibinfo
  {journal} {J. Acoust. Soc. Am}\ }\textbf {\bibinfo {volume} {103}},\ \bibinfo
  {pages} {2971} (\bibinfo {year} {1998})}\BibitemShut {NoStop}%
\bibitem [{\citenamefont {Wunenburger}\ \emph {et~al.}(2015)\citenamefont
  {Wunenburger}, \citenamefont {Lozano},\ and\ \citenamefont
  {Brasselet}}]{wunenburger2015acoustic}%
  \BibitemOpen
  \bibfield  {author} {\bibinfo {author} {\bibfnamefont {R.}~\bibnamefont
  {Wunenburger}}, \bibinfo {author} {\bibfnamefont {J.~I.~V.}\ \bibnamefont
  {Lozano}},\ and\ \bibinfo {author} {\bibfnamefont {E.}~\bibnamefont
  {Brasselet}},\ }\bibfield  {title} {\bibinfo {title} {Acoustic orbital
  angular momentum transfer to matter by chiral scattering},\ }\href@noop {}
  {\bibfield  {journal} {\bibinfo  {journal} {New Journal of Physics}\ }\textbf
  {\bibinfo {volume} {17}},\ \bibinfo {pages} {103022} (\bibinfo {year}
  {2015})}\BibitemShut {NoStop}%
\bibitem [{\citenamefont {Hong}\ \emph {et~al.}(2015)\citenamefont {Hong},
  \citenamefont {Zhang},\ and\ \citenamefont
  {Drinkwater}}]{hong2015observation}%
  \BibitemOpen
  \bibfield  {author} {\bibinfo {author} {\bibfnamefont {Z.~Y.}~\bibnamefont
  {Hong}}, \bibinfo {author} {\bibfnamefont {J.}~\bibnamefont {Zhang}},\ and\
  \bibinfo {author} {\bibfnamefont {B.~W.}\ \bibnamefont {Drinkwater}},\
  }\bibfield  {title} {\bibinfo {title} {Observation of orbital angular
  momentum transfer from bessel-shaped acoustic vortices to diphasic
  liquid-microparticle mixtures},\ }\href@noop {} {\bibfield  {journal}
  {\bibinfo  {journal} {Physical review letters}\ }\textbf {\bibinfo {volume}
  {114}},\ \bibinfo {pages} {214301} (\bibinfo {year} {2015})}\BibitemShut
  {NoStop}%
\bibitem [{\citenamefont {Bliokh}\ and\ \citenamefont
  {Nori}(2019)}]{bliokh2019spin}%
  \BibitemOpen
  \bibfield  {author} {\bibinfo {author} {\bibfnamefont {K.~Y.}\ \bibnamefont
  {Bliokh}}\ and\ \bibinfo {author} {\bibfnamefont {F.}~\bibnamefont {Nori}},\
  }\bibfield  {title} {\bibinfo {title} {Spin and orbital angular momenta of
  acoustic beams},\ }\href@noop {} {\bibfield  {journal} {\bibinfo  {journal}
  {Physical Review B}\ }\textbf {\bibinfo {volume} {99}},\ \bibinfo {pages}
  {174310} (\bibinfo {year} {2019})}\BibitemShut {NoStop}%
\bibitem [{\citenamefont {Toftul}\ \emph {et~al.}(2019)\citenamefont {Toftul},
  \citenamefont {Bliokh}, \citenamefont {Petrov},\ and\ \citenamefont
  {Nori}}]{toftul2019acoustic}%
  \BibitemOpen
  \bibfield  {author} {\bibinfo {author} {\bibfnamefont {I.~D.}~\bibnamefont
  {Toftul}}, \bibinfo {author} {\bibfnamefont {K.~Y.}~\bibnamefont {Bliokh}},
  \bibinfo {author} {\bibfnamefont {M.~I.}\ \bibnamefont {Petrov}},\ and\
  \bibinfo {author} {\bibfnamefont {F.}~\bibnamefont {Nori}},\ }\bibfield
  {title} {\bibinfo {title} {Acoustic radiation force and torque on small
  particles as measures of the canonical momentum and spin densities},\
  }\href@noop {} {\bibfield  {journal} {\bibinfo  {journal} {Physical review
  letters}\ }\textbf {\bibinfo {volume} {123}},\ \bibinfo {pages} {183901}
  (\bibinfo {year} {2019})}\BibitemShut {NoStop}%
\bibitem [{\citenamefont {Oemrawsingh}\ \emph {et~al.}(2004)\citenamefont
  {Oemrawsingh}, \citenamefont {Van~Houwelingen}, \citenamefont {Eliel},
  \citenamefont {Woerdman}, \citenamefont {Verstegen}, \citenamefont
  {Kloosterboer} \emph {et~al.}}]{oemrawsingh2004production}%
  \BibitemOpen
  \bibfield  {author} {\bibinfo {author} {\bibfnamefont {S.}~\bibnamefont
  {Oemrawsingh}}, \bibinfo {author} {\bibfnamefont {J.}~\bibnamefont
  {Van~Houwelingen}}, \bibinfo {author} {\bibfnamefont {E.}~\bibnamefont
  {Eliel}}, \bibinfo {author} {\bibfnamefont {J.}~\bibnamefont {Woerdman}},
  \bibinfo {author} {\bibfnamefont {E.}~\bibnamefont {Verstegen}}, \bibinfo
  {author} {\bibfnamefont {J.}~\bibnamefont {Kloosterboer}}, \emph {et~al.},\
  }\bibfield  {title} {\bibinfo {title} {Production and characterization of
  spiral phase plates for optical wavelengths},\ }\href@noop {} {\bibfield
  {journal} {\bibinfo  {journal} {Applied Optics}\ }\textbf {\bibinfo {volume}
  {43}},\ \bibinfo {pages} {688} (\bibinfo {year} {2004})}\BibitemShut
  {NoStop}%
\bibitem [{\citenamefont {Cromb}\ \emph {et~al.}(2020)\citenamefont {Cromb},
  \citenamefont {Gibson}, \citenamefont {Toninelli}, \citenamefont {Padgett},
  \citenamefont {Wright},\ and\ \citenamefont
  {Faccio}}]{cromb2020amplification}%
  \BibitemOpen
  \bibfield  {author} {\bibinfo {author} {\bibfnamefont {M.}~\bibnamefont
  {Cromb}}, \bibinfo {author} {\bibfnamefont {G.~M.}\ \bibnamefont {Gibson}},
  \bibinfo {author} {\bibfnamefont {E.}~\bibnamefont {Toninelli}}, \bibinfo
  {author} {\bibfnamefont {M.~J.}\ \bibnamefont {Padgett}}, \bibinfo {author}
  {\bibfnamefont {E.~M.}\ \bibnamefont {Wright}},\ and\ \bibinfo {author}
  {\bibfnamefont {D.}~\bibnamefont {Faccio}},\ }\bibfield  {title} {\bibinfo
  {title} {Amplification of waves from a rotating body},\ }\href@noop {}
  {\bibfield  {journal} {\bibinfo  {journal} {Nature Physics}\ }\textbf
  {\bibinfo {volume} {16}},\ \bibinfo {pages} {1069} (\bibinfo {year}
  {2020})}\BibitemShut {NoStop}%
\end{thebibliography}

\providecommand{\noopsort}[1]{}\providecommand{\singleletter}[1]{#1}%
%
%

\clearpage
\setcounter{figure}{0}
\renewcommand{\thefigure}{S\arabic{figure}}
\section{Supplemental Material}
\maketitle
\section{Laguerre-Gaussian Beams}
Here we briefly recap the theory of Laguerre-Gaussain beams in the paraxial approximation of the wave equation to form the comparison with the full 3D treatment of the waves in elastic pipes. 

Allen et al \cite{allen1992orbital} demonstrate that Laguerre-Gaussian beams carry a well defined orbital angular momentum by first considering the angular momentum density associated with the transverse EM field, given as \cite{jackson1999classical}
\begin{equation}
    \boldsymbol{M} = \epsilon_{0} \boldsymbol{r} \times \langle\boldsymbol{E} \times \boldsymbol{B}\rangle
    \label{eq:M}
\end{equation}
where $\boldsymbol{E}, \boldsymbol{B}$ are the electric and magnetic fields, $\epsilon_0$ is the vacuum permittivity and $\boldsymbol{r}$ the position vector. In the Lorentz gauge the laser field of a linearly polarised beam, propagating in the $z$-direction, can be written in terms of the vector potential 
\begin{equation}
    \boldsymbol{A}(\boldsymbol{r}) = u(\boldsymbol{r})e^{-ikz},
\end{equation}
where $\boldsymbol{A}$ can refer to either the magnetic or electric field vector. The complex scalar function $u(\boldsymbol{r})$ satisfies the paraxial wave equation
\begin{equation}
\nabla_{\perp}^{2} + 2ik\frac{\partial u}{\partial z} = 0,
\end{equation}
where $\nabla_{\perp}^{2}$ is the transverse part of the Laplacian. Physically this asserts that $\partial_{z}u$ varies slowly in $z$, a property satisfies by most laser beams. In cylindrical polars the complex amplitude $u_{pl}(r,\theta,z)$ defines the Laguerre-Gaussian modes
\begin{align}
    \begin{split}
        u_{pl}(r,\theta,z) &= \frac{C_{pl}}{\sqrt{1+\left(\frac{z}{z_R}\right)^2}}\left(\frac{r\sqrt{2}}{w(z)}\right)^{l}L_{p}^{l}\left(\frac{2r^2}{w^2(z)}\right) \\
        &\times \exp\left(\frac{-r^2}{w^2(z)}\right)\exp\left(\frac{-ikr^2z}{2\left(z^2+z_R^2\right)}\right)\\\ &\times \exp\left(-il\phi\right)\exp\left(i(2p+l+1)\tan^{-1}\left(\frac{z}{z_R}\right)\right),
    \end{split}
\end{align}
where $z_R$ is the Rayleigh range, $w(z)$ the radius of curvature of the beam, $L_{p}^{l}$ an associated Laguerre polynomial, $C_{pl}$ a constant. The beam waist is centred at $z = 0$. Utilising these modes \eqref{eq:M} yields an OAM of $l\hbar$ per photon. 

\section{Guided waves in pipes}
\subsection{Analytical Solutions}
Considering transverse and longitudinal wave motion, the elastic displacement vector $\bxi$ can be written in terms of the curl-less dilatational scalar potential, $\Phi$, and the divergence-less equivoluminal vector shear potential, $\bPsi$, such that
\begin{align}
\begin{split}
    \bxi &= \bxi_{L} + \bxi_{T} \\
    &= \nabla\Phi + \nabla\times\bPsi.
    \end{split}
\end{align}
Using this, the equations of elastodynamics reduce to two wave equations for compressional and shear waves such that
\begin{align}
\begin{split}
    \nabla^2\Phi &= c_p^{-2}\ddot{\Phi}, \quad c_p = \sqrt{\frac{\lambda + 2\mu}{\rho}} \\
    \nabla^2\bPsi &= c_s^{-2}\ddot{\bPsi}, \quad c_s = \sqrt{\frac{\mu}{\rho}},
\end{split}
\label{eq:waveeqns2}
\end{align}
with $c_p$ and $c_s$ being the compressional and shear bulk wavespeeds respectively. Lam\'{e}'s first and second parameters take the form
\begin{align}
    \begin{split}
        \lambda &= \frac{E\nu}{(1+\nu)(1-2\nu)} \\ 
\mu &= \frac{E}{2(1+\nu)},
    \end{split}
\end{align}
for Young's modulus $E$ and Poisson's ratio $\nu$.
Given the cylindrically symmetric nature of the pipe system we pose the solutions
\begin{equation}
\begin{split}
    \Phi &= \phi(r)\exp(i(m\theta + k_{z}z - \omega t)), \\
    \Psi_{r} &= \psi_{r}(r)\exp(i(m\theta + k_{z}z - \omega t)), \\
    \Psi_{\theta} &= \psi_{\theta}(r)\exp(i(m\theta + k_{z}z - \omega t)), \\
    \Psi_{z} &= \psi_{z}(r)\exp(i(m\theta + k_{z}z - \omega t)), \\
\end{split}
\label{eq:Expansions}
\end{equation} 
where $m$ is the azimuthal order of the wave travelling along the $k_z$ direction (i.e. along the pipe axis). Substituting the ansatz \eqref{eq:Expansions} into \eqref{eq:waveeqns2}, results in Bessel's and modified Bessel's equations for the functions describing the radial components, for example 
\begin{equation}
    r^2\phi^{\prime\prime} +r\phi^{\prime} +\left[\left(\frac{\omega^2}{c_p^2} - k^2\right)r^2 - m^2\right]\phi = 0.
    \label{eq:Bessel}
\end{equation}
The full set of equations are solved by  \cite{gazis1959a,gazis1959b,rose2014ultrasonic}:
\begin{align}
    \phi(r) &= AZ_{m}(\alpha_{1}r) + BW_{m}(\alpha_{1}r) \\
    \begin{split}
    h_{1} &\equiv \frac{1}{2}(i\psi_{r}(r) -\psi_{\theta}(r)) \\&= A_{1}Z_{m+1}(\beta_{1}r) + B_{1}W_{m+1}(\beta_{1}r)
    \end{split}\\
    \begin{split}
    h_{2} &\equiv \frac{1}{2}(i\psi_{r}(r) +\psi_{\theta}(r)) \\&= A_{2}Z_{m-1}(\beta_{1}r) + B_{2}W_{m-1}(\beta_{1}r)\\
    \end{split} \\
    h_{3} &\equiv \psi_{z}(r) = A_{3}Z_{m}(\beta_{1}r) + B_{3}W_{m}(\beta_{1}r)
\end{align}
where 
\begin{align}
    \begin{split}
        Z_{m}(\alpha_{1}r) = \begin{cases}
        J_m(\alpha_{1}r), &  \frac{\omega^2}{c_p^2}-k_z^2 \geq 0, \\
        I_m(\alpha_{1}r), &  \frac{\omega^2}{c_p^2}-k_z^2 < 0,
        \end{cases} \\
        W_{m}(\alpha_{1}r) = \begin{cases}
        Y_m(\alpha_{1}r), &  \frac{\omega^2}{c_p^2}-k_z^2 \geq 0, \\
        K_m(\alpha_{1}r), &  \frac{\omega^2}{c_p^2}-k_z^2 < 0,
        \end{cases}
        \label{eq:Bessel1}
    \end{split} \\ 
    \begin{split}
        Z_{m}(\beta_{1}r) = \begin{cases}
        J_m(\beta_{1}r), &  \frac{\omega^2}{c_s^2}-k_z^2 \geq 0, \\
        I_m(\beta_{1}r), &  \frac{\omega^2}{c_s^2}-k_z^2 < 0,
        \end{cases} \\
        W_{m}(\beta_{1}r) = \begin{cases}
        Y_m(\beta_{1}r), &  \frac{\omega^2}{c_s^2}-k_z^2 \geq 0, \\
        K_m(\beta_{1}r), &  \frac{\omega^2}{c_s^2}-k_z^2 < 0,
        \end{cases}
        \label{eq:Bessel2}
    \end{split}
\end{align}
with $\alpha_{1} \equiv |\alpha| = |\frac{\omega^2}{c_p^2} - k_z^2|$ and $\beta_{1} \equiv |\beta| = |\frac{\omega^2}{c_s^2} - k_z^2|$ and $J_m(x), Y_m(x)$ are Bessel functions of the first and second kind respectively, with $I_{m}(x), K_{m}(x)$ the corresponding modified Bessel functions.

Traction free boundary conditions on the inner and outer radii, $r_a$ and $r_b$ such that
\begin{equation}
    \sigma_{rr} = \sigma_{r\theta} = \sigma_{rz} = 0\big\rvert_{r_{a,b}},
    \label{eq:tractionfree}
\end{equation}
and the infinitely long cylinder gauge
\begin{equation}
    \nabla\bm{\cdot}\bPsi = 0
    \label{eq:gauge}
\end{equation}
are used to construct a linear set of equations to solve for the coefficients $A_{i}, B_{i}$, from which the field components can be retrieved
\begin{align}
    \xi_{r} &= \frac{\partial\Phi}{\partial r} + \frac{im}{r}\Psi_{z} - ik_z\Psi_{\theta}, \\
    \xi_{\theta} &= \frac{im}{r}\Phi + ik_z\Psi_{r} - \frac{\partial\Psi_{z}}{\partial r}, \\
    \xi_{z} &= ik_{z}\Phi + \frac{\Psi_{\theta}}{r} + \frac{\partial\Psi_{\theta}}{\partial r} -\frac{im}{r}\Psi_{r}.
\end{align}

As an alternative to cumbersome, traditional root-finding schemes to recover the coefficients $A_{i}, B_{i}$ or to Finite Element Methods (FEM) \cite{marzani2008SAFE}, we instead opt to utilise a Spectral Collocation Method (SCM), based on that by Adamou and Craster \cite{adamou2004spectral} and now widely used, for instance, in DISPERSE a commercial package for finding dispersion curves \cite{pavlakovic1997disperse}. SCM is versatile, accurate and has found application in elastic waveguide problems, a typical more recent example being for anisotropic waveguides \cite{quintanilla2015anisotropic}.

\subsection{Spectral Collocation Method}
Assuming time-harmonicity, as we do, the governing equations \eqref{eq:waveeqns2} can be written as the eigen-problems
\begin{equation}
    \begin{split}
        &\mathcal{L}\Phi = -\frac{\omega^2}{c_p^2}\Phi, \\ \\
        &\left(\left(\mathcal{L} - \frac{1}{r^2}\right)\Psi_{r} - \frac{2im}{r^2}\Psi_{\theta}\right)\boldsymbol{\hat{e}}_{r} = -\frac{\omega^2}{c_s^2}\Psi_{r}\boldsymbol{\hat{e}}_{r}, \\ \\
        &\left(\frac{2im}{r^2}\Psi_{r} + \left(\mathcal{L} - \frac{1}{r^2}\right)\Psi_{\theta} \right)\boldsymbol{\hat{e}}_{\theta} = -\frac{\omega^2}{c_s^2}\Psi_{\theta}\boldsymbol{\hat{e}}_{\theta}, \\ \\
        &\mathcal{L}\Psi_{z} = -\frac{\omega^2}{c_s^2}\Psi_{z},
    \end{split}
    \label{eq:operatoreqn}
\end{equation}
with
\begin{equation}
    \mathcal{L} = \left(\frac{d^2}{dr^2} + \frac{1}{r}\frac{d}{dr} - \left(\frac{m^2}{r^2} + k_{z}^2 \right) \right).
\end{equation}
The spectral collocation method employed throughout rests on representing this differential operator as a differentiation matrix. We use Chebyshev differentiation matrices, namely that multiplication by the $n^{th}$-order Chebyshev differentiation matrix $D^{(n)}$
transforms a vector of data at $N$ Chebyshev interpolation points into approximate derivatives at those points. Following the steps outlined in \cite{adamou2004spectral} permits \eqref{eq:operatoreqn} to be written as a matrix eigenvalue equation:
\begin{equation}
    {P}\bxi = -\omega^2Q\bxi,
\end{equation}
with $\bxi = \left(\Phi(r_i),\Psi_{r}(r_{i}),\Psi_{\theta}(r_{i}),\Psi_{z}(r_{i})\right)^{T}$ where $i = 1,\hdots,N$, and the matrices $P$ and $Q$ encode the boundary conditions. The rapid, spectrally accurate solutions obtained with this method allow the calculation of the dispersion curves of an infinite hollow elastic cylinder with ease. We adopt this method, notably in calculation of the dispersion curves in Fig.~2(a) in the main text, and to numerically evaluate $\log|\frac{2\mathcal{J}_{Lz}}{\omega\rho c_p^2}|$ in Fig.2(b), whereby we demonstrate that the elastic OAM associated with the dilatational potential of flexural modes in pipes is well defined. The pipe considered is of Aluminium with inner diameter $40~\si{\milli\meter}$, thickness $h = 10~\si{\milli\meter}$ and has density $\rho = 2710~\si{\kilo\gram\meter^{-3}}$, Young's Modulus $E = 70~\si{\giga\pascal}$, and Poisson's ratio $\nu = 0.33$.

In Fig. 2(a) we cross-validate the SCM results with dispersion curves obtained from FEM calculations, using the commercial FEM software COMSOL Multiphysics \textsuperscript{\textregistered}\cite{comsol}. To do this we take a section of the pipe of (arbitrary) length $a = 10$ mm, and apply Floquet-Bloch periodic boundary conditions to the faces along the length of the pipe; the periodicity of this pipe section then represents an infinitely long pipe. The ensuing eigenvalue problem is solved for wave-vectors up to $k_z = \pi/a$, highlighting an advantage of the SCM: in the FEM solutions for $k_z > \pi/a$ are `band-folded' due to the artificial periodicity introduced by the pipe section, and as such require post-processing. This is not required in the SCM.

\begin{figure*}
    \centering
    \includegraphics[width = 0.95\textwidth]{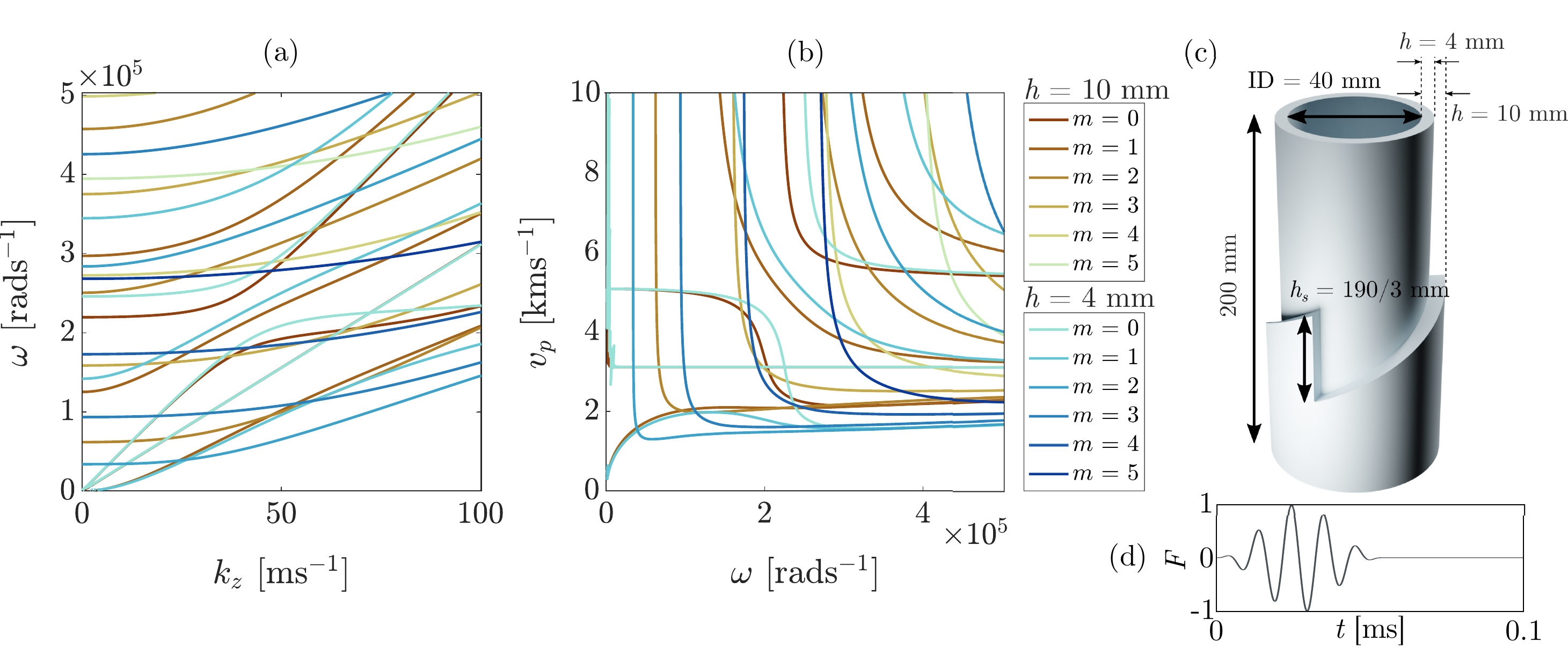}
    \caption{(a,b) Dispersion curves for infinite pipe of inner diameter $ID = 40~\si{\milli\meter}$ and thicknesses $h_1 = 4~\si{\milli\meter}$ (blue) and $h_2 = 10~\si{\milli\meter}$ (red). (c) Schematic of eSPP used in FEM simulations. (d) Hanning window excitation (normalised force) applied on top boundary of the pipe (marked by arrows in Fig. 3 in the main text).}
    \label{fig:disps}
\end{figure*}

\section{Orbital Angular momentum density: Calculation details, transverse and hybrid contributions}

We then exploit the standard definition of the material derivative in cylindrical polars, such that $\left[\left(\boldsymbol{A}\bm{\cdot}\nabla \right)\boldsymbol{B}\right] \bm{\cdot}\bm{\hat{\theta}}$
\begin{align}
\begin{split}
    = \bigg(A_{r}\frac{\partial B_{\theta}}{\partial r} &+ \frac{A_{\theta}}{r}\frac{\partial B_\theta}{\partial\theta} +A_z\frac{\partial B_{\theta}}{\partial z} + \frac{A_{\theta}B_{r}}{r} \bigg).
    \end{split}
\end{align}
To evaluate $\boldsymbol{M}_{L}^{o}\bm{\cdot}\bm{\hat{z}} = r\boldsymbol{p}_{L}^{o}\bm{\cdot}\bm{\hat{\theta}}$, we are required to calculate
\begin{align}
    \begin{split}
    \mathfrak{Im}&\left[\left(\bxi_{L}^{*}\bm{\cdot}\nabla\right)\bxi_{L} \right]\bm{\cdot}\bm{\hat{\theta}} = \\
    \mathfrak{Im}&\left[m\left(\frac{i|\phi^{\prime}|^2}{r} + \frac{im^2|\phi|^2}{r^3}+\frac{ik^2|\phi|^2}{r}-\frac{i}{r^2}\left(\phi\phi^{*\prime} + \phi^{\prime}\phi^{*} \right) \right) \right] \\
    &= \frac{m}{r}\left(|\phi^{\prime}|^2 +\left(\frac{m^2}{r^2}+k^2 \right)|\phi|^2 -\frac{2}{r}\mathfrak{Re}\left[\phi\phi^{*\prime}\right] \right),
    \end{split}
\end{align}
and as such
\begin{equation}
\begin{split}
   = \frac{m\omega\rho c_p^2}{2}\left(|\phi^{\prime}|^2 +\left(\frac{m^2}{r^2}+k^2 \right)|\phi|^2 -\frac{2}{r}\mathfrak{Re}\left[\phi\phi^{*\prime}\right] \right).
    \end{split}
\end{equation}
Using Bessel's equation \eqref{eq:Bessel}, this can be re-written to the final expression presented in the main text.

The contributions to the elastic OAM density from the transverse and hybrid parts are such that 
\begin{align}
    \begin{split}
        &\boldsymbol{M}_{T}^{o}\bm{\cdot}\bm{\hat{z}} \equiv r\boldsymbol{p}_{T}^{o}\bm{\cdot}\bm{\hat{\theta}} = \frac{\omega\rho r}{2}c_s^2\mathfrak{Im}\left[\left(\bxi^{*}_{T}\bm{\cdot}\nabla\right)\bxi_{T}\right]\bm{\cdot}\bm{\hat{\theta}}, \\
        &\boldsymbol{M}_{H}^{o}\bm{\cdot}\bm{\hat{z}} \equiv r\boldsymbol{p}_{H}^{o}\bm{\cdot}\bm{\hat{\theta}} \\ &= \frac{\omega\rho r}{2}\left( c_p^2\mathfrak{Im}\left[\left(\bxi^{*}_{T}\bm{\cdot}\nabla\right)\bxi_{L}\right] + c_s^2\mathfrak{Im}\left[\left(\bxi^{*}_{L}\bm{\cdot}\nabla\right)\bxi_{T}\right]\right)\bm{\cdot}\bm{\hat{\theta}}.
        \label{eq:adv}
    \end{split}
\end{align}
We show here that these components do not possess a well-defined OAM as additional terms are present which are not proportional to the azimuthal index $m$. Working in cylindrical polar coordinates gives
\begin{align}
    \begin{split}
    \bxi_{L} &\equiv \nabla\Phi = \left(\phi^{\prime},\frac{im\phi}{r},ik\phi \right)e^{i(m\theta + kz - \omega t)} \\ 
    \bxi_{T} &\equiv \nabla\times\bPsi = \left(\frac{im\psi_{z}}{r}-ik\psi_{\theta} \right)e^{i(m\theta + kz - \omega t)}\boldsymbol{\hat{r}} \\
    &+ \left(ik\psi_{r} - \psi_{z}^{\prime} \right)e^{i(m\theta + kz - \omega t)}\boldsymbol{\hat{\theta}} \\
    &+\left(\frac{\psi_{\theta}}{r} + \psi_{\theta}^{\prime} - \frac{im\psi_{r}}{r} \right)e^{i(m\theta + kz - \omega t)}\boldsymbol{\hat{z}},
    \end{split}
\end{align}
We then substitute these expressions, and their complex conjugates, into the advective terms in \eqref{eq:adv}. Considering first the term $\mathfrak{Im}\left[\left(\bxi^{*}_{T}\bm{\cdot}\nabla\right)\bxi_{T}\right]\bm{\cdot}\bm{\hat{\theta}}$, we find, after some algebra, expressions such as 
\begin{equation}
    -\frac{k^2}{r}\left(\psi_{\theta}^{*}\psi_{r} + \psi_{\theta}\psi_{r}^{*} \right),
\end{equation}
which, of course, is purely real. As such terms such as these vanish when we take the imaginary part. Continuing in this fashion we eventually arrive at the expression
\begin{align}
    \begin{split}
    &\mathfrak{Im}\left[\left(\bxi^{*}_{T}\bm{\cdot}\nabla\right)\bxi_{T}\right]\bm{\cdot}\bm{\hat{\theta}} = -\mathfrak{Im}\left[k\partial_{r}\left( \psi_{\theta}^{*}\left(i\psi_{z}^{\prime}+k\psi_{r} \right)\right) \right] \\&+ \frac{m}{r}\mathfrak{Im}\left[\partial_{r}\left(k\psi_{z}^{*}\psi_{r}  + i \psi_{z}^{\prime}\psi_{z}^{*}\right) + \frac{1}{r}\left(k\psi_{r}^{*} - i\psi_{z}^{\prime *} \right)\psi_{z}\right],
    \end{split}
\end{align}
where we denote $\partial_{r} = \partial/\partial r$. This is clearly not solely $\propto m$ and hence is not well-defined in the same sense as the contribution from the compressional potential.

Similarly we find
\begin{align}
    \begin{split}
        &\mathfrak{Im}\left[\left(\bxi_{T}^{*}\bm{\cdot}\nabla\right)\bxi_{L} \right]\bm{\cdot}\bm{\hat{\theta}} = -\frac{1}{r}\mathfrak{Im}\left[\left( ik\psi_{r}^{*} - \psi_{z}^{\prime *}\right)\phi^{\prime}
        \right] \\
        &+ \frac{m}{r}\mathfrak{Im}\left[\partial_{r}\left(\left(\frac{m\psi_{z}^{*}}{r} - k\psi_{\theta}^{*} \right)\phi\right) \right],
    \end{split}
\end{align}
and
\begin{align}
    \begin{split}
        &\mathfrak{Im}\left[\left(\bxi_{L}^{*}\bm{\cdot}\nabla\right)\bxi_{T} \right]\bm{\cdot}\bm{\hat{\theta}} \\&= \mathfrak{Im}\left[\phi^{\prime *}\left(ik\psi_{r}^{\prime} - \psi_{z}^{\prime\prime} \right) + k^2\phi^{*}\left(ik\psi_{r} -\psi_{z}^{\prime} \right) \right] \\
        &+ \frac{m}{r}\mathfrak{Im}\left[\phi^{*}\left(\frac{ikm\psi_{r}}{r} -\frac{k\psi_{\theta}}{r} -m\partial_{r}\left(\frac{\psi_{z}}{r}\right) \right)\right].
    \end{split}
\end{align}
Therefore both the transverse and hybrid orbital angular momentum components ($\bm{\mathcal{J}}_{T}\bm{\cdot}\bm{\hat{z}} = \int \bm{M}_{T}^{o}\bm{\cdot}\bm{\hat{z}}d\mathbf{r}$ and $\bm{\mathcal{J}}_{H}\bm{\cdot}\bm{\hat{z}} = \int \bm{M}_{H}^{o}\bm{\cdot}\bm{\hat{z}}d\mathbf{r}$ respectively) do not carry a well defined OAM proportional to $m$.

\section{The Elastic Spiral Phase Pipe \& FEM Modelling}
To demonstrate the transfer of elastic OAM in a coupled solid-fluid system, efficient excitation of pure flexural modes in elastic pipes is required. Here we outline the theory behind the elastic spiral phase pipe \cite{chaplain2021elastic} by first considering its optical analogues.

In optics, wave-fields with inclined phase fronts, such as LG beams are vortex beams characterised by a phase singularity at the beam centre with locally vanishing intensity. These beams carry a topological charge defined as \cite{oemrawsingh2004production}
\begin{equation}
    \mathcal{Q} = \frac{1}{2\pi}\oint d\chi,
\end{equation}
where $\chi$ is the phase of the field. Such laser modes can be excited by optical spiral phase plates. These are refractive devices with an azimuthally-dependent height variation parameterised by a circular helicoid. The height of the spiral step, $h_s$, is chosen such that the optical path difference experienced by a traversing wave introduces a phase retardation that results in helical phase fronts.

The elastic spiral phase pipe is a recent translation of this device to elastic systems; it consists of hollowed circular helicoid (as shown in Fig.~1 in the main text) whose step profile height $h_s$ is calculated by relating the phase speeds of the incident mode and desired converted mode through a relative refractive index, $\tilde{n} = c_f/c_i$. In the examples used throughout the main text we consider an incident $L(0,2)$ mode with phase speed $c_i$ (and therefore wavelength $\lambda_i = 2\pi/k_i$) that is mode converted into a flexural $F(3,2)$ after the device with phase speed $c_f$. The step height takes the form
\begin{equation}
    h_s = \frac{2\pi\mathcal{Q}}{k_i(\tilde{n}-1)}.
    \label{eq:hs}
\end{equation}
\begin{figure}
    \centering
    \includegraphics[width = 0.4\textwidth]{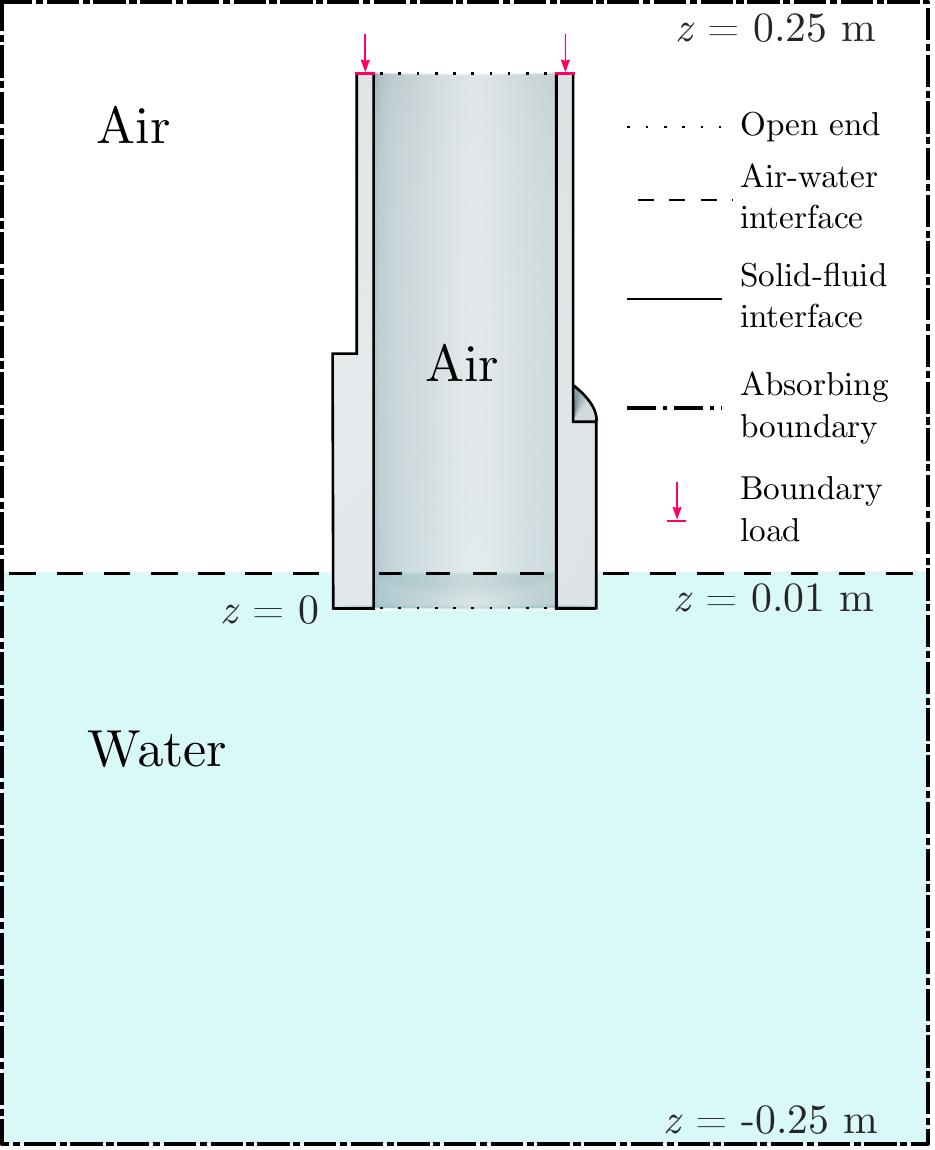}
    \caption{Cross section of FEM simulation domain, showing fluid and solid regions with interfaces. The corresponding boundary conditions are implemented in COMSOL Multiphysics\textsuperscript{\textregistered}.}
    \label{fig:FEM_X}
\end{figure}

For the demonstration of elastic OAM transfer, we utilise the experimentally verified eSPP of \cite{chaplain2021elastic}, with the same properties as the pipe considered in Fig. 2 in the main text. We choose a frequency of operation of $62~\si{\kilo\hertz}$ and form an eSPP by removing $6~\si{\milli\meter}$ of Aluminium (which can be milled experimentally) with an azimuthally varying profile determine by the step profile \eqref{eq:hs}. This is calculated by once again leveraging the SCM to evaluate the dispersion curves for the two regions of the pipe with thicknesses $h_1 = 4~\si{\milli\meter}$ and $h_2 = 10~\si{\milli\meter}$. Two representations of the dispersion curves are shown in Fig.~\ref{fig:disps}(a,b); those corresponding to $h_1$ are shown in blue hues whilst $h_2$ are shown in red. At the operating frequency the step height is then determined to be $h_s = 190~\si{\milli\meter}$; in the FEM simulations presented in the main text this height is partitioned over 3-spiral steps, as shown in Fig.~\ref{fig:disps}(c).
\begin{figure}
    \centering
    \includegraphics[width = 0.375\textwidth]{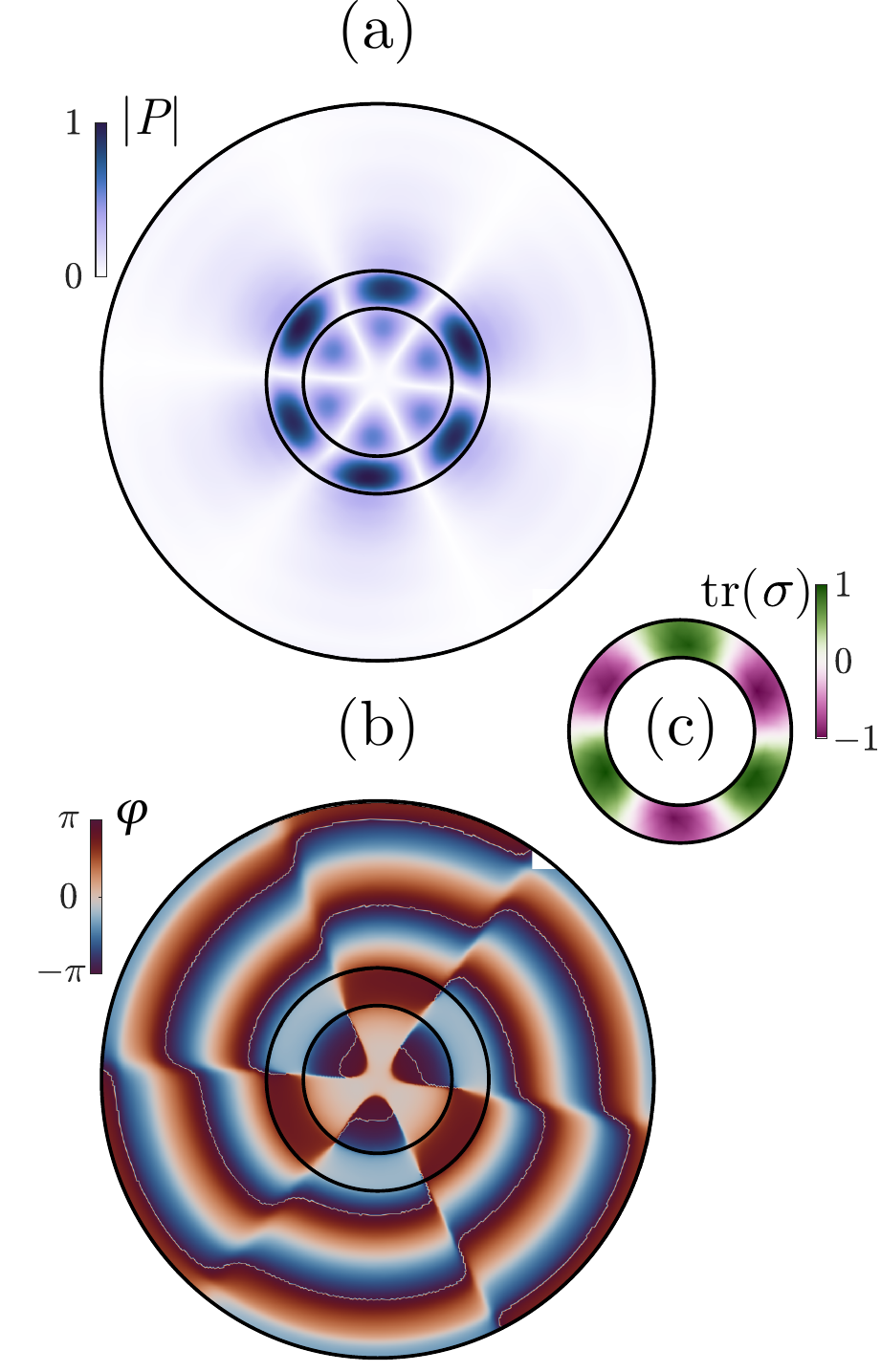}
    \caption{FEM frequency domain simulations: (a) Normalised absolute pressure field. (b) Phase of Pressure field, $\varphi$. (c) Trace of stress tensor, $\tr(\sigma)$, i.e. the pressure distribution at the end of the pipe. In all three cases the results are shown at the base of the submerged end of the pipe at $z = 0$. The excitation is a continuous sinusoidal boundary load, at the boundary shown in Fig. S2, at a frequency of $62$ kHz.}
    \label{fig:FEM_freq}
\end{figure}
The time domain simulation presented in the main text results from FEM simulation performed using COMSOL Multiphysics\textsuperscript{\textregistered} \cite{comsol}, employing multiphysics coupling between the acoustics and structural mechanics module such that the acoustic-structure boundary is modelled to include the fluid load on the structure, and the structural acceleration experienced by the fluid. We implement a boundary longitudinal forcing on the top end of the pipe, with absorbing boundary conditions (perfectly matched layers) on the exterior of the fluid domain. A cross-section of the domain, with all boundary conditions, is shown in Fig.~\ref{fig:FEM_X}. The pipe is partially submerged $1 $ cm in water (density $\rho$ = $1000$ kgm$^{-3}$), with a uniform water level inside and outside the pipe. Air fills and surrounds the rest of the pipe. A 5-cycle Hanning window centred on $62~\si{\kilo\hertz}$ excites an axisymmetric $L(0,2)$ mode in the thin region of the pipe. After traversing the eSPP region this is endowed with a helical phase profile and is mode converted into the $F(3,2)$ mode with high efficiency, which carries elastic OAM. The compressional potential then couples with the pressure field in the fluid at the submerged end of the pipe, exciting spiraling acoustic waves within the fluid, demonstrating the transfer of OAM.

The generation of the $F(3,2)$ wave in the time domain simulation is validated by comparisons to both the dispersion curves obtained from the SCM and FEM (for the case of the pipe in-vacuo), along with frequency-domain simulations (shown in Fig.~\ref{fig:FEM_freq}) and experimental corroboration from Ref.~\cite{chaplain2021elastic}. The validation of the full multiphysics simulation is verified through standard time-stepping convergence methods \cite{comsol}. In Fig. S3 we additionally show the spiralling phase distribution of the acoustic field in the fluid, as well as the trace of the stress tensor, $\tr(\sigma)$, since the normal stress match the pressure at the walls of the pipe

\end{document}